\documentclass[11pt]{article}
\usepackage{cite}
\usepackage{type1cm}
\usepackage[dvipdfm]{graphicx}
\usepackage{amsmath,amssymb,bm,amsthm}

\def\beq{\begin{equation}}
\def\eeq{\end{equation}}
\def\nbeq{\begin{equation*}}
\def\neeq{\end{equation*}}

\def\<{\langle}
\def\>{\rangle}
\def\rt#1{\sqrt{\mathstrut #1}}
\def\Tr{{\rm Tr}}

\renewcommand{\d}{\partial}

\makeatletter
\newcommand{\subscripts}[3]{%
  \@mathmeasure\z@\displaystyle{#2}%
  \global\setbox\@ne\vbox to\ht\z@{}\dp\@ne\dp\z@
  \setbox\tw@\box\@ne
  \@mathmeasure4\displaystyle{\copy\tw@_{#1}}%
  \@mathmeasure6\displaystyle{{#2}_{#3}}%
  \dimen@-\wd6 \advance\dimen@\wd4 \advance\dimen@\wd\z@
  \hbox to\dimen@{}\mathop{\kern-\dimen@\box4\box6}%
}
\makeatother

\begin{document}

\title{Exactness of the mean-field dynamics in optical cavity systems}
\author{Takashi Mori \\
{\it
Department of Physics, Graduate School of Science,} \\
{\it The University of Tokyo, Bunkyo-ku, Tokyo 113-0033, Japan
}}
\maketitle

\begin{abstract}
Validity of the mean-field approach to open system dynamics in the optical cavity system is examined.
It is rigorously shown that the mean-field approach is justified in the thermodynamic limit.
The result is applicable to nonequilibrium situations, e.g. the thermal reservoirs may have different temperatures,
and the system may be subject to a time-dependent external field.
The result of this work will lead to further studies on macroscopic open quantum systems.
\end{abstract}

\section{Introduction}

Quantum dynamics in many body systems is a very important topic for many branches of physics.
In particular, recent studies in the field of statistical physics have focused on dynamical problems such as thermalization
~\cite{Polkovnikov_review2011,Deutsch1991,Srednicki1994,Goldstein_PRE2010,Tasaki2010}, 
phase transitions induced by the parameter quench~\cite{Kibble1976,Zurek1985,Zurek2005,Sciolla-Biroli2011},
and nonequilibrium phase transitions~\cite{Kastner2011,Diehl2010,Knap2011}.
Theoretically, it is difficult to precisely describe the time evolution of a many body system.
Although recent works succeeded to treat the exact time evolution and observe relaxation processes 
in some integrable systems~\cite{Faribault2009,Mossel2010,Sato_PRL2012},
we should be content with some approximate treatments in general.

One of the important approximate approaches is the mean-field (MF) approximation for not only equilibrium statics but also nonequilibrium dynamics.
In this approximation, the $N$-body state vector or the $N$-body density matrix is approximated by the product state.
Interestingly, the MF approach predicts some remarkable results 
such as the absence of thermalization and purely dynamical phase transitions in some isolated quantum systems~\cite{Sciolla-Biroli2011,Kastner2011}.
In open quantum systems, nonequilibrium phase transitions have also been studied by the MF approach
~\cite{Hepp-Lieb1973,Drummond-Walls1980,Drummond1981}; see also Ref.~\cite{Shirai2012} for a recent result.

Some of these predictions might be the peculiarity of the MF theory and not universal in general.
However, importantly, there are several models in which the MF treatment becomes exact in some ideal limit
~\cite{Spohn_review1980,Erdoes2007,Schlein_lecture2008,Merkli-Berman2012}.
Thus the MF theory is completely reliable as long as such an ideal limit is considered to be realized.
For example, in a quantum spin system with a global coupling (the infinite-range interaction),
it is known that the quantum dynamics is exactly described by the MF dynamical equation (or Hartree equation)~\cite{Spohn_review1980}.
Another example is an $N$-body bosonic system interacting via a two-body potential with scattering length $a$,
in which the time-dependent Gross-Pitaevskii equation, which is regarded as the MF theory of the dynamics of the Bose-Einstein condensate,
becomes exact in the limit of $N\rightarrow\infty$ with a fixed value of $Na$~\cite{Erdoes2007}.
Because of the existence of such ideal limits,
the MF approximation takes a special position among other approximate approaches.

Therefore, we should not regard the MF approach as a mere crude approximation.
Rather, it is important to clarify the condition under which the MF approach is justified and
extend the possibility of experimental realization of the remarkable predictions of the MF approach.

This paper is mainly devoted to investigating the quantum dynamics of 
a system consisting of $N$ two-level atoms interacting with a single quantized boson mode in an optical cavity.
This system is described by the Dicke model~\cite{Dicke1954}.
We consider its extension,
that is, we allow each element of the system, i.e. single cavity mode of the photons and $N$ two-level atoms, 
to be attached to its own environment.
Each environment is a Hamilton system, and may or may not be a large system.
If one environment has a continuous energy spectrum, which means that the environment is an infinitely large system, it plays the role of a dissipative thermal bath.

The MF approach of this model predicts interesting dynamical or nonequilibrium phase transitions.
In the isolated Dicke model (there is no environment), it was argued that 
dynamical phase transitions occur by the MF approach~\cite{Sciolla-Biroli2011}.
In open systems (the Dicke model interacting with dissipative environments), under the time-periodic driving field,
the MF dynamics combined with the method of the Born-Markov quantum master equation~\cite{Breuer_text} 
predicts some nonequilibrium phase transitions including optical bistabilities~\cite{Drummond-Walls1980,Drummond1981} 
which were observed in experiments~\cite{Gripp1996}.
It has been believed that the MF approach is justified in the limit of $N\rightarrow\infty$
because in this model $N$ two-level atoms interact with a common cavity mode of photons, 
and it is expected that each atom feels only the MF produced by the other atoms via radiation and absorption of cavity photons.
In this situation, the correlation between two atoms or between an atom and the cavity photons will be not so important.
However, there has not been rigorous proof of this expectation yet.

In this paper, we rigorously prove that the MF approach becomes exact in the limit of an infinite number of two-level atoms
in the sense that the expectation value of an observable belonging to the restricted set ${\cal B}$, which will be specified in Sec.~\ref{sec:observable}, 
at an arbitrary time $t$ is exactly equal to that calculated by the MF theory.
The system may be in contact with thermal reservoirs and may be subject to a time-dependent external field,
and the result does not depend on whether the dynamics of the system of interest is Markovian or not.
Thus the result is very general.
The use of the MF theory in the above mentioned previous works on this model is justified.
The restriction of this work, which should be removed in future works, 
is that the Hamiltonian must be linear with respect to cavity photons and the Bose filed attaching to them (``0-subsystem'' in Sec.~\ref{sec:model}).

The strategy of the proof is based on the comparison of the Bogoliubov-Born-Green-Kirkwood-Yvon (BBGKY) hierarchy generated by the exact equations
with the one generated by the MF equations.
We will see that they become indistinguishable in the thermodynamic limit.
This strategy itself is applicable to simpler models, i.e. infinite-range spin models, so that we will first apply this strategy to those simple models, 
and then proceed to the extended Dicke model, in which the proof is much more complicated.

The organization of this paper is as follows.
In Sec.~\ref{sec:spin}, we introduce the proof of the fact that the MF theory is justified in infinite-range spin models.
This result is already known~\cite{Spohn_review1980}, but we review it because it is the simplest case where the MF approach is justified rigorously.
In Sec.~\ref{sec:model}, the model considered in this work is explained.
In Sec.~\ref{sec:preliminaries}, some preliminaries are covered.
We introduce the two notions important later, i.e. the $Q$-representation of the density matrix and the restriction of the class of physical observables.
In Sec.~\ref{sec:statement}, the main result of this paper is stated, and its proof is summarized.
The detailed evaluation of some quantities necessary for the proof is given in Appendices~\ref{appendix:bound_B} and \ref{appendix:bound_B_prime}.
In Sec.~\ref{sec:conclusion}, we summarize the result of this work and discuss some future problems.

\section{Exactness of the mean-field dynamics in infinite-range spin models}
\label{sec:spin}

Before studying the optical cavity system, let us analyze the simplest case where the MF dynamics becomes exact in the thermodynamic limit.
We consider an $N$ spin system with infinitely long-range interactions.
The Hilbert space for $i$th spin is denoted by ${\cal H}_i$.
Let $h_i=h$ be a bounded self-adjoint operator acting on ${\cal H}_i$,
and let $V_{ij}=V$ ($i\neq j$) also be a bounded self-adjoint operator acting on ${\cal H}_i\otimes{\cal H}_j$.
We put $V_{ii}=0$.
The Hamiltonian of the spin system with infinite-range two-body interactions is generally written as
\beq
H=\sum_{i=1}^Nh_i+\frac{1}{N}\sum_{i,j=1}^NV_{ij}.
\eeq

We introduce the $p$-norm of an operator $A$ acting on $\bigotimes_{i=1}^k{\cal H}_i$ as
\beq
\|A\|_p^{(k)}:=\left(\Tr_{1,2,\dots,k} |A|^p\right)^{1/p}.
\eeq
In particular, the operator norm is defined as 
\beq
\|A\|_{\infty}:=\sup_{\psi\in\bigotimes_{i=1}^k{\cal H}_i}\frac{\<\psi||A||\psi\>}{\<\psi|\psi\>}.
\eeq
The operator $A$ is said to be bounded if $\|A\|_{\infty}<+\infty$.

The $N$-spin density matrix at time $t$ is denoted by $\rho_{N,t}$, and its $k$-marginal reduced density matrix by $\gamma_{N,t}^{(k)}$ which is defined as
\beq
\gamma_{N,t}^{(k)}:=\Tr_{k+1,k+2,\dots,N}\rho_{N,t}.
\eeq
The initial condition is assumed to satisfy
\beq
\lim_{N\rightarrow\infty}\left|\gamma_{N,0}^{(k)}-\gamma_0^{\otimes k}\right|_1^{(k)}=0 \quad \forall k\in\mathbb{N},
\label{eq:d_initial}
\eeq
that is, reduced density matrices are initially factorized.
Here, $\gamma_0$ is a single spin density matrix
and $$\gamma_0^{\otimes k}:=\underbrace{\gamma_0\otimes\gamma_0\otimes\dots\otimes\gamma_0}_k.$$
It is remarked that both the Hamiltonian and the initial state are site-symmetric, and hence the density matrix remains site-symmetric during the time evolution.

The MF theory assumes that reduced density matrices are always factorized,
\beq
\gamma_{N,t}^{(k)}\approx\gamma_{{\rm MF},t}^{(k)}=\gamma_t^{\otimes k} \quad \forall k\in\mathbb{N}.
\eeq
The single spin density matrix $\gamma_t$ obeys the Hartree equation,
\begin{align}
\frac{d}{dt}\gamma_t&=-i{\cal L}_1^{(0)}\gamma_t-i\Tr_2{\cal L}_{12}^{(V)}\gamma_t\otimes\gamma_t
\nonumber \\
&=-i[h_1,\gamma_t]+\Tr_2[V_{12},\gamma_t\otimes\gamma_t].
\end{align}
The Liouville operators are defined as ${\cal L}_i^{(0)}(\cdot):=[h_i,(\cdot)]$ and ${\cal L}_{ij}^{(V)}(\cdot):=[V_{ij},(\cdot)]$.

The statement of exactness of the MF dynamics is that if the initial state satisfies Eq.~(\ref{eq:d_initial}),
\beq
\lim_{N\rightarrow\infty}\left\|\gamma_{N,t}^{(k)}-\gamma_t^{\otimes k}\right\|_1^{(k)}=0
\label{eq:d_converge}
\eeq
for any fixed $k\in\mathbb{N}$ and $t>0$.\footnote
{Exactness of the MF dynamics does {\it not} mean $\rho_{N,t}=\gamma_t^{\otimes N}$.
Actually, it does not hold even in infinite-range spin models.}

From Eq.~(\ref{eq:d_converge}) it is immediately verified, by using the inequality $\|AB\|_1^{(k)}\leq\|A\|_{\infty}\|B\|_1^{(k)}$, that
\beq
\lim_{N\rightarrow\infty}\left|\left\<{\cal O}^{(k)}\right\>_{N,t}-\left\<{\cal O}^{(k)}\right\>_{{\rm MF},t}\right|=0,
\eeq
for any {\it bounded} operator ${\cal O}^{(k)}$ acting on $\bigotimes_{i=1}^k{\cal H}_i$.
Here $\<\cdot\>_{N,t}:=\Tr(\cdot)\rho_{N,t}$ and $\<\cdot\>_{{\rm MF},t}:=\Tr_{1,2,\dots,k}(\cdot)\gamma_{{\rm MF},t}^{(k)}$.

We follow Ref.~\cite{Schlein_lecture2008} for the proof of Eq.~(\ref{eq:d_converge}). 
We start with the Liouville equation
\beq
\frac{d}{dt}\rho_{N,t}=-i\left(\sum_{i=1}^N{\cal L}_i^{(0)}+\frac{1}{N}\sum_{i,j=1}^N{\cal L}_{ij}^{(V)}\right)\rho_{N,t}.
\eeq
By tracing out over $\bigotimes_{i=k+1}^N{\cal H}_k$, we obtain the hierarchical equations (BBGKY hierarchy) 
for $\{\gamma_{N,t}^{(k)}\}$~\cite{Schlein_lecture2008},
\beq
\frac{d}{dt}\gamma_{N,t}^{(k)}=-i\sum_{i=1}^k{\cal L}_i^{(0)}\gamma_{N,t}^{(k)}+{\cal W}^{(k)}\gamma_{N,t}^{(k+1)}
+\frac{1}{N}{\cal V}^{(k)}\gamma_{N,t}^{(k)}-\frac{k}{N}{\cal W}^{(k)}\gamma_{N,t}^{(k+1)}.
\label{eq:d_BBGKY}
\eeq
The super-operators ${\cal V}^{(k)}$ and ${\cal W}^{(k)}$ are defined by
\beq
\left\{
\begin{aligned}
{\cal V}^{(k)}\gamma_{N,t}^{(k)}&:=-i\sum_{i,j=1}^k{\cal L}_{ij}^{(V)}\gamma_{N,t}^{(k)}, \\
{\cal W}^{(k)}\gamma_{N,t}^{(k+1)}&:=-i\Tr_{k+1}\sum_{i=1}^k{\cal L}_{i,k+1}^{(V)}\gamma_{N,t}^{(k+1)}.
\end{aligned}
\right.
\eeq
The MF solution $\gamma_{{\rm MF},t}^{(k)}=\gamma_t^{\otimes k}$ satisfies the equation
\beq
\frac{d}{dt}\gamma_{{\rm MF},t}^{(k)}=-i\sum_{i=1}^k{\cal L}_i^{(0)}\gamma_{{\rm MF},t}^{(k)}+{\cal W}^{(k)}\gamma_{{\rm MF},t}^{(k+1)},
\label{eq:d_MF_BBGKY}
\eeq
which is obtained by formally taking the limit of $N\rightarrow\infty$ in Eq.~(\ref{eq:d_BBGKY}).

It should be remarked that the fact that Eq.~(\ref{eq:d_MF_BBGKY}) is obtained from Eq.~(\ref{eq:d_BBGKY}) 
by formally taking the limit of $N\rightarrow\infty$ alone
does not ensure that the MF dynamics is exact in this limit.
Because Eq.~(\ref{eq:d_BBGKY}) forms a coupled chain of $N$ equations of motion, 
the terms proportional to $1/N$ in Eq.~(\ref{eq:d_BBGKY}) might be amplified in the chain and have nonnegligible contribution.
In particular, for a very large $k\lesssim N$, the last two terms on the RHS of Eq.~(\ref{eq:d_BBGKY}) are no longer small.
We must prove that these terms actually do not influence the dynamics of $\gamma_{N,t}^{(k)}$.

The next step is to construct the {\it Duhamel series}~\cite{Schlein_lecture2008} by formally integrating Eq.~(\ref{eq:d_BBGKY}).
We define
$$U_t^{(k)}:=\exp\left[-i\sum_{i=1}^k{\cal L}_i^{(0)}t\right].$$
Then we obtain
\begin{align}
\gamma_{N,t}=&U_t^{(k)}\gamma_{N,0}^{(k)}+\int_0^tdt_1U_{t-t_1}^{(k)}{\cal W}^{(k)}\gamma_{N,t_1}^{(k+1)}
\\
&+\frac{1}{N}\int_0^tdt_1U_{t-t_1}^{(k)}{\cal V}^{(k)}\gamma_{N,t_1}^{(k)}
-\frac{k}{N}\int_0^tdt_1U_{t-t_1}^{(k)}{\cal W}^{(k)}\gamma_{N,t_1}^{(k+1)}.
\end{align}
We substitute this expression iteratively into the terms on the RHS not proportional to $1/N$ (the second term of the RHS in the above equation),
and repeat this procedure $L$ times, then we obtain the following so called the Duhamel series expansion,
\begin{align}
\gamma_{N,t}^{(k)}=&\sum_{l=0}^{L-1}\int_0^tdt_1\int_0^{t_1}dt_2\dots\int_0^{t_{l-1}}dt_l
U_{t-t_1}^{(k)}{\cal W}^{(k)}U_{t_1-t_2}^{(k+1)}{\cal W}^{(k+1)}
\dots {\cal W}^{k+l-1}U_{t_l}^{(k+l)}\gamma_{N,0}^{(k+l)}
\nonumber \\
&+\int_0^tdt_1\int_0^{t_1}dt_2\dots\int_0^{t_{L-1}}dt_LU_{t-t_1}^{(k)}{\cal W}^{(k)}U_{t_1-t_2}^{(k+1)}{\cal W}^{(k+1)}
\dots U_{t_{L-1}-t_L}^{(k+L-1)}{\cal W}^{(k+L-1)}\gamma_{N,t_L}^{(k+L)}
\nonumber \\
&+\frac{1}{N}\sum_{l=1}^L\int_0^tdt_1\int_0^{t_1}dt_2\dots\int_0^{t_{l-1}}dt_l
U_{t-t_1}^{(k)}{\cal W}^{(k)}U_{t_1-t_2}^{(k+1)}{\cal W}^{(k+1)}
\nonumber \\ 
&\qquad\qquad\qquad\qquad\qquad\qquad\dots U_{t_{l-1}-t_l}^{(k+l-1)}{\cal V}^{(k+l-1)}\gamma_{N,t_l}^{(k+l-1)}
\nonumber \\
&-\sum_{l=1}^L\frac{k+l-1}{N}\int_0^tdt_1\int_0^{t_1}dt_2\dots\int_0^{t_{l-1}}dt_l
U_{t-t_1}^{(k)}{\cal W}^{(k)}U_{t_1-t_2}^{(k+1)}{\cal W}^{(k+1)}
\nonumber \\
&\qquad\qquad\qquad\qquad\qquad\qquad\dots U_{t_{l-1}-t_l}^{(k+l-1)}{\cal W}^{(k+l-1)}\gamma_{N,t_l}^{(k+l)}.
\label{eq:d_Duhamel}
\end{align}

The Duhamel series of the MF density matrix is written as
\begin{align}
\gamma_{{\rm MF},t}^{(k)}=\sum_{l=0}^{L-1}\int_0^tdt_1\dots\int_0^{t_{l-1}}dt_l
U_{t-t_1}^{(k)}{\cal W}^{(k)}U_{t_1-t_2}^{(k+1)}{\cal W}^{(k+1)}
\dots {\cal W}^{(k+l-1)}U_{t_l}^{(k+l)}\gamma_{{\rm MF},0}^{(k+l)}
\nonumber \\
+\int_0^tdt_1\int_0^{t_1}dt_2\dots\int_0^{t_{L-1}}dt_LU_{t-t_1}^{(k)}{\cal W}^{(k)}U_{t_1-t_2}^{(k+1)}{\cal W}^{(k+1)}
\dots U_{t_{L-1}-t_L}^{(k+L-1)}{\cal W}^{(k+L-1)}\gamma_{{\rm MF},t_L}^{(k+L)}.
\label{eq:d_MF_Duhamel}
\end{align}

From these expressions, we shall show $\|\gamma_{N,t}^{(k)}-\gamma_{{\rm MF},t}^{(k)}\|_1^{(k)}\rightarrow 0$ in the limit of $N\rightarrow\infty$.
The key point is that we choose $L$ so that $1\ll L\ll N$, which is realized by taking the limit of $N\rightarrow\infty$ first and then the limit of $L\rightarrow\infty$.
From Eqs.~(\ref{eq:d_Duhamel}) and (\ref{eq:d_MF_Duhamel}), we obtain
\beq
\left\|\gamma_{N,t}^{(k)}-\gamma_{\rm MF}^{(k)}\right\|_1^{(k)}\leq A_1+A_2+A_3+A_4,
\eeq
where $A_i$ ($i=1,2,3,4$) are given by
\begin{align*}
A_1=&\sum_{l=0}^{L-1}\int_0^tdt_1\int_0^{t_1}dt_2\dots\int_0^{t_{l-1}}dt_l
\left\|U_{t-t_1}^{(k)}{\cal W}^{(k)}U_{t_1-t_2}^{(k+1)}{\cal W}^{(k+1)} \right.
\\
&\left.\qquad\qquad\qquad\qquad\qquad
\dots {\cal W}^{k+l-1}U_{t_l}^{(k+l)}\left(\gamma_{N,0}^{(k+l)}-\gamma_{{\rm MF},0}^{(k+l)}\right)\right\|_1^{(k)},
\\
A_2=&\int_0^tdt_1\int_0^{t_1}dt_2\dots\int_0^{t_{L-1}}dt_L\left\|U_{t-t_1}^{(k)}{\cal W}^{(k)}U_{t_1-t_2}^{(k+1)}{\cal W}^{(k+1)}\right.
\\
&\left.\qquad\qquad\qquad\qquad
\dots U_{t_{L-1}-t_L}^{(k+L-1)}{\cal W}^{(k+L-1)}\left(\gamma_{N,t_L}^{(k+L)}-\gamma_{{\rm MF},t_L}^{(k+L)}\right)\right\|_1^{(k)},
\\
A_3=&\frac{1}{N}\sum_{l=1}^L\int_0^tdt_1\int_0^{t_1}dt_2\dots\int_0^{t_{l-1}}dt_l
\left\|U_{t-t_1}^{(k)}{\cal W}^{(k)}U_{t_1-t_2}^{(k+1)}{\cal W}^{(k+1)}\right.
\\ 
&\left.\qquad\qquad\qquad\qquad\qquad\qquad\dots U_{t_{l-1}-t_l}^{(k+l-1)}{\cal V}^{(k+l-1)}\gamma_{N,t_l}^{(k+l-1)}\right\|_1^{(k)},
\\
A_4=&\sum_{l=1}^L\frac{k+l-1}{N}\int_0^tdt_1\int_0^{t_1}dt_2\dots\int_0^{t_{l-1}}dt_l
\left\|U_{t-t_1}^{(k)}{\cal W}^{(k)}U_{t_1-t_2}^{(k+1)}{\cal W}^{(k+1)}\right.
\\
&\left.\qquad\qquad\qquad\qquad\qquad\qquad\dots U_{t_{l-1}-t_l}^{(k+l-1)}{\cal W}^{(k+l-1)}\gamma_{N,t_l}^{(k+l)}\right\|_1^{(k)}.
\end{align*}

Now we evaluate $A_i$.
Since $U_t^{(k)}$ is a unitary operator, it does not change the norm, $\|U_t^{(k)}(\cdot)\|_1^{(k)}=\|(\cdot)\|_1^{(k)}$.
We further use the inequality $\|AB\|_1^{(k)}\leq\|A\|_{\infty}\|B\|_1^{(k)}$.
As a result, $A_1$ is evaluated, by recalling the definitions of ${\cal V}^{(k)}$ and ${\cal W}^{(k)}$, as
$$
A_1\leq\sum_{l=0}^{L-1}\frac{t^l}{l!}(2\|V\|_{\infty})^l\frac{(k+l-1)!}{(k-1)!}\left\|\gamma_{N,0}^{(k+l)}-\gamma_{{\rm MF},0}^{(k+l)}\right\|_1^{(k+l)}.
$$
By the assumption for the initial condition, Eq.~(\ref{eq:d_initial}), $A_1\rightarrow 0$ as $N\rightarrow\infty$ at a fixed $L$.
Similarly, the upper bound of $A_2$ becomes
\begin{align*}
A_2&\leq\frac{t^L}{L!}(2\|V\|_{\infty})^L\frac{(k+L-1)!}{(k-1)!}\left\|\gamma_{N,t_L}^{(k+l)}-\gamma_{{\rm MF},t_L}^{(k+l)}\right\|_1^{(k+l)}
\\
&\leq\frac{(k+L-1)!}{L!(k-1)!}(2\|V\|_{\infty}t)^L\times 2.
\end{align*}
In the last inequality, we used $\left\|\gamma_{N,t_L}^{(k+l)}-\gamma_{{\rm MF},t_L}^{(k+l)}\right\|_1^{(k+l)}\leq 2$.
Since $(k+L-1)!/[L!(k-1)!]\leq 2^{k+L-1}$, we have
$$A_2\leq 2^k(4\|V\|_{\infty}t)^L.$$
If we restrict the time to $t\leq t_0:=1/(8\|V\|_{\infty})$ (this restriction will be removed later), we have $A_2\leq 2^{k-L}$.
If we take the limit of $L\rightarrow\infty$ after $N\rightarrow\infty$,
both $A_1$ and $A_2$ converge to zero.

Similarly, we can show that $A_3$ and $A_4$ also converge to zero in the same limit.
The upper bound of $A_3$ is evaluated as
\begin{align}
A_3&\leq\frac{1}{N}\sum_{l=1}^{L-1}\frac{t^l}{l!}(2\|V\|_{\infty})^l(k+l-1)\frac{(k+l-1)!}{(k-1)!}
\nonumber \\
&=\frac{1}{N}\sum_{l=1}^{L-1}(2\|V\|_{\infty}t)^l(k+l-1)\frac{(k+l-1)!}{l!(k-1)!}
\nonumber \\
&\leq\frac{1}{N}\sum_{l=1}^{L-1}2^{k-1}(k+l-1)(4\|V\|_{\infty}t)^l.
\label{eq:d_A3_upper}
\end{align}
If we take the limit of $N\rightarrow\infty$ first, obviously $A_3$ goes to zero.
$A_4$ is also bounded from above by Eq.~(\ref{eq:d_A3_upper}).
Therefore, it has been proven that $\lim_{L\rightarrow\infty}\lim_{N\rightarrow\infty}A_i=0$
for the time interval $0\leq t\leq t_0$.

The restriction of $t_0$ is not essential.
Since $\left\{\gamma_{N,t_0}^{(k)}\right\}$ satisfies the assumption of Eq.~(\ref{eq:d_initial}),
we can start with $\left\{\gamma_{N,t_0}^{(k)}\right\}$ at time $t_0$ as a new initial condition.
By applying the same argument as above, it is concluded that Eq.~(\ref{eq:d_converge}) holds in the time interval $0\leq t\leq 2t_0$.
Repeating this argument, we can extend $t_0\rightarrow\infty$, and Eq.~(\ref{eq:d_converge}) is proved for an arbitrary time $t\geq 0$.

The above proof relies on the boundedness of the operators.
In optical systems, however, the above proof cannot be used as it is since the creation and annihilation operators of bosons are unbounded.
We construct the proof for such a situation in the following sections.

\section{Model}
\label{sec:model}

We consider the Schr\"odinger dynamics of the generalized Dicke model given by
\begin{align}
H(t)=\sum_{j=0}^N\left(H_{{\rm S}_j}(t)+H_{{\rm B}_j}+H_{{\rm I}_j}\right)+V, 
\label{eq:Dicke}
\\
\left\{
\begin{aligned}
H_{{\rm S}_0}(t)&=\omega_pa^{\dagger}a+\rt{N}\xi(t)(a+a^{\dagger}), \\
H_{{\rm S}_j}&=\omega_a\sum_{i=1}^NS_j^z \qquad (j=1,2,\dots, N), \\
H_{{\rm B}_0}&=\sum_r\omega_rb_r^{\dagger}b_r, \\
H_{{\rm I}_0}&=\sum_r\lambda_r(b_ra^{\dagger}+b_r^{\dagger}a), \\
V&=\frac{g}{\rt{N}}\sum_{j=1}^N(aX_j^{\dagger}+a^{\dagger}X_j).
\end{aligned}
\right.
\label{eq:model}
\end{align}
In $H_{{\rm S}_j}(t)$, the index $j=0$ corresponds to the single mode of cavity photons driven by the external field $\xi(t)$, 
and $j=1,2,\dots,N$ corresponds to the ensemble of $N$ two-level atoms.
Cavity photons ($H_{{\rm S}_0}$) interact with the free Bose field $H_{{\rm B}_0}$ through the coupling $H_{{\rm I}_0}$.
Each two-level atom $H_{{\rm S}_j}$, $j=1,2,\dots N$, interacts with an arbitrary Hamilton system $H_{{\rm B}_j}$
through an arbitrary interaction Hamiltonian $H_{{\rm I}_j}$.
We call the system of the Hamiltonian $H_{{\rm B}_j}$ ``$j$-environment'' ($j=0,1,\dots,N$),
and call the composite system described by the Hamiltonian $H_{{\rm S}_j}(t)+H_{{\rm B}_j}+H_{{\rm I}_j}$ ``$j$-subsystem''.
We assume that the operators $H_{{\rm B}_j}+H_{{\rm I}_j}$ are identical for $j=1,2,\dots,N$;
the Hamiltonian $H(t)$ is symmetric under the exchange of two indices $i$ and $j$ ($i$, $j=1,2,\dots,N$).

The last term of Eq.~(\ref{eq:Dicke}) represents the interaction between the cavity photons and the ensemble of atoms.
If we choose $X_j=S_j^x$ $(S_j^-)$, the Hamiltonian $\sum_{j=0}^NH_{{\rm S}_j}(t)+V$ is called the Dicke model 
(Tavis-Cummings model~\cite{Tavis-Cummings1968}), with a driving force $\xi$.
Therefore, we can regard Eq.~(\ref{eq:Dicke}) as an extension of the Dicke-like model; environmental systems are attached to it.

If the $j$-environment has a continuous spectrum, then this environment acts as a thermal reservoir in contact with $H_{{\rm S}_j}$.
We can obtain the Hamiltonian with a continuous spectrum as a limiting case of the discrete spectrum.
It is necessary to properly choose the coupling constants $\{\lambda_r\}$ between cavity photons and 0-environment in order to have a well-defined limit.
We introduce a parameter $\Lambda$ so that the number of eigenmodes with the frequencies between $\omega$ and $\omega+d\omega$
is given by $\Lambda D(\omega)d\omega$.
Thus the limit of $\Lambda\rightarrow\infty$ corresponds to the limit of a continuum of 0-environmental modes.
We assume that $\lambda_r\sim\lambda(\omega_r)/\rt{\Lambda}.$
In this case, if we consider the limit of the continuous spectrum of the 0-environment, $\sum_r\rightarrow\Lambda\int_0^{\infty}d\omega D(\omega)$,
and therefore, $\sum_r|\lambda_r|^2\rightarrow\int_0^{\infty}d\omega D(\omega)\lambda(\omega)^2=:\int_0^{\infty}d\omega J(\omega)$
does not depend on $\Lambda$.
Here, $J(\omega)$ is called the spectral density~\cite{Weiss_text}.
We assume that 
\beq
\sum_r|\lambda_r|^2<+\infty
\eeq
in order not to diverge the effect of the 0-environment on the dynamics of cavity photons.
In general, for the proof of the justification of the MF treatment, $\Lambda$ is arbitrary (it can even depend on $N$).
When $\Lambda\ll N$, the effect of the 0-environment on the dynamics of the whole system is negligible.
When $\Lambda\sim N$, the 0-environment affects the dynamics of the total system.
When $\Lambda\gg N$, the 0-environment behaves as a thermal reservoir, and causes the dissipation.

\subsection*{Mean-field dynamics}

The density matrix of the whole system, $\rho_{N,t}$, is defined on the Hilbert space $\bigotimes_{i=0}^N{\cal H}_i$,
where ${\cal H}_j$ ($j=0,1,2,\dots,N$) is the Hilbert space of the $j$-subsystem.
The MF theory assumes that the $k$-marginal reduced density matrix $\gamma_{N,t}^{(k)}=\Tr_{k+1,\dots,N}\rho_{{\rm MF},t}$ is in the product form,
\beq
\gamma_{N,t}^{(k)}\approx\gamma_{{\rm MF},t}^{(k)}=(\rho_{\rm p})_t\otimes(\rho_{\rm a})_t^{\otimes k}.
\eeq
$(\rho_{\rm p})_t$ is the density matrix for the 0-subsystem.
Similarly, $(\rho_{\rm a})_t$ is a common density matrix for the $k$-subsystem ($k=1,2,\dots,N$).
The density matrices $(\rho_{\rm p})_t$ and $(\rho_{\rm a})_t$ obey the following Hartree equations:
\begin{align}
\frac{d}{dt}(\rho_{\rm p})_t=&-i\left[\omega_pa^{\dagger}a+\rt{N}\xi(t)(a+a^{\dagger})+H_{{\rm B}_0}+H_{{\rm I}_0}
\right.\nonumber \\
&\left.+\rt{N}g\left\{[{\rm Tr}X_1(\rho_{\rm a})_t]a^{\dagger}+\rt{N}g[{\rm Tr}X_1^{\dagger}(\rho_{\rm a})_t]a\right\},(\rho_{\rm p})_t\right],
\label{eq:MF_p}
\\
\frac{d}{dt}(\rho_{\rm a})_t=&-i\left[\omega_aS_1^z+H_{{\rm B}_1}+H_{{\rm I}_1}
+\frac{g}{\rt{N}}\left\{[{\rm Tr}a(\rho_{\rm p})_t]X_1^{\dagger}+[{\rm Tr}a^{\dagger}(\rho_{\rm p})_t]X_1\right\},(\rho_{\rm a})_t\right].
\label{eq:MF_a}
\end{align}
The apparent $N$-dependence can be removed by an appropriate scaling in the coherent state representation, see Sec.~\ref{sec:Q-rep}.

It is stressed that the MF theory here corresponds to neglecting the correlation between $i$- and $j$-subsystems.
In general, we cannot neglect the correlation between the system of interest ($H_{{\rm S}_j}$) and environments ($H_{{\rm I}_j}$).

Rich nonequilibrium phase transitions of the model~(\ref{eq:model}) have been studied with the help of the MF theory,
including the optical bistability~\cite{Drummond-Walls1980,Drummond1981} and 
the spontaneously symmetry-broken phases~\cite{Shirai2012} in the regime of the strong coupling (large $g$) and the strong field (large $\xi(t)=\xi\cos(\Omega t)$).
Thus it is physically important to establish the validity of the MF theory in this model.

\section{Preliminaries}
\label{sec:preliminaries}

Although we cannot justify the MF theory straightforwardly in the same way as we did in the spin systems,
we follow essentially the same course.
We derive the BBGKY hierarchy for the reduced density matrices, construct the Duhamel series expansion, and evaluate each term.
However, because of the unboundedness of boson operators
we cannot use the trace norm, and the statement itself must be modified.
Some preliminaries are necessary before presenting the main result of this work.

\subsection{Duhamel series expansion in the coherent state representation}
\label{sec:Q-rep}

Since the density matrix $\rho_{N,t}$ obeys the Liouville equation
\beq
\frac{d}{dt}\rho_{N,t}=-i[H(t),\rho_{N,t}]=:-i{\cal L}\rho_{N,t},
\eeq
we obtain the following chain of equations of motion for $\left\{\gamma_{N,t}^{(k)}\right\}$,
\begin{align}
\frac{d}{dt}\gamma_{N,t}^{(k)}=&-i\left[\sum_{j=0}^k\left(H_{{\rm S}_j}+H_{{\rm B}_j}+H_{{\rm I}_j}\right),
\gamma_{N,t}^{(k)}\right]
\nonumber \\
&-i\left[\frac{g}{\rt{N}}\left(a\sum_{j=1}^kX_j^{\dagger}+a^{\dagger}\sum_{j=1}^kX_j\right),\gamma_{N,t}^{(k)}\right]
\nonumber \\
&-i(N-k){\rm Tr}_{k+1}\left[\frac{g}{\rt{N}}\left(aX_{k+1}^{\dagger}+a^{\dagger}X_{k+1}\right),\gamma_{N,t}^{(k+1)}\right].
\end{align}

Because it is hard to see which terms are important in the limit of $N\rightarrow\infty$,
and because the relation to the MF theory is not obvious in this form,
let us introduce the Husimi ``Q-representation'' of the reduced density matrix~\cite{Husimi1940}:
\beq
Q_{N,t}^{(k)}(\alpha,\{\beta_r\})=\left(\frac{N}{\pi}\right)^{{\cal N}+1}
\subscripts{0}{\left\<\rt{N}\alpha;\{\rt{N}\beta_r\}\left|\gamma_{N,t}^{(k)}\right|\rt{N}\alpha;\{\rt{N}\beta_r\}\right\>}{0}.
\label{eq:Q}
\eeq
Here ${\cal N}$ is the number of modes of 0-environment (${\cal N}=\sum_r1\simeq\Lambda\int_0^{\infty}D(\omega)d\omega$).
The vector $\left|\rt{N}\alpha,\{\rt{N}\beta_r\}\right\>_0$ denotes the coherent state on the Hilbert space ${\cal H}_0$, and it satisfies
\begin{align}
a\left|\rt{N}\alpha,\{\rt{N}\beta_r\}\right\>_0&=\rt{N}\alpha\left|\rt{N}\alpha,\{\rt{N}\beta_r\}\right\>_0,
\\
b_r\left|\rt{N}\alpha,\{\rt{N}\beta_r\}\right\>_0&=\rt{N}\beta_r\left|\rt{N}\alpha,\{\rt{N}\beta_r\}\right\>_0.
\end{align}
Because $\gamma_{N,t}^{(k)}$ is a matrix on $\bigotimes_{j=0}^k{\cal H}_j$ and 
$\left|\rt{N}\alpha,\{\rt{N}\beta_r\}\right\>_0$ is a vector on ${\cal H}_0$,
$Q_{N,t}^{(k)}$ is still a matrix on $\bigotimes_{j=1}^k{\cal H}_j$, that is, the Hilbert space of $k$ atoms and their environments.
The average in Eq.~(\ref{eq:Q}) is taken only over the Hilbert space of the 0-subsystem, ${\cal H}_0$.

$Q_{N,t}^{(k)}$ has the following properties:
\begin{align*}
&\text{ (i) } \quad Q_{N,t}^{(k)}\geq 0, \\
&\text{ (ii) } \quad \int d^2\alpha\left(\prod_{r}\int d^2\beta_r\right){\rm Tr}_{1,2,\dots,k}Q_{N,t}^{(k)}=1, \\
&\text{ (iii) } \quad \Tr_{0,1,\dots,k}{\cal O}^{(k)}
\left(\frac{a}{\rt{N}},\frac{a^{\dagger}}{\rt{N}},\left\{\frac{b_r}{\rt{N}},\frac{b_r^{\dagger}}{\rt{N}}\right\}\right)
\gamma_{N,t}^{(k)} \\
&\qquad\qquad =\int d^2\alpha\left(\prod_{r}\int d^2\beta_r\right)
{\rm Tr}_{1,2,\dots,k}{\cal O}^{(k)}(\alpha,\alpha^*,\{\beta_r,\beta_r^*\})Q_{N,t}^{(k)},
\end{align*}
where ${\cal O}^{(k)}$ is an arbitrary {\it anti-normal ordered} operator acting on $\bigotimes_{j=0}^k{\cal H}_j$;
namely, all the annihilation operators are in the left of all the creation operators.
From the above properties, we can obtain the expectation value of any observable ${\cal O}^{(k)}$ from $Q_{N,t}^{(k)}$.

An advantage to using $Q_{N,t}^{(k)}$ instead of $\gamma_{N,t}^{(k)}$ is that it becomes easy to see which terms are important and 
which terms are likely to be negligible in the limit of $N\rightarrow\infty$.
Indeed, the time evolution equation of $Q_{N,t}^{(k)}$ is given by
\beq
\frac{d}{dt}Q_{N,t}^{(k)}=-i({\cal L}_{\rm p}(t)+{\cal L}_{\rm a}^{(k)}(\alpha))Q_{N,t}^{(k)}
+\left(1-\frac{k}{N}\right){\cal W}^{(k)}Q_{N,t}^{(k+1)}+\frac{1}{N}{\cal V}^{(k)}Q_{N,t}^{(k)},
\label{eq:dQdt}
\eeq
where
\begin{align}
-i{\cal L}_{\rm p}(t)Q_{N,t}^{(k)}:=&i\frac{\d}{\d\alpha}\left[\left(\omega_p\alpha+\xi(t)
+\sum_r\lambda_r\beta_r\right)Q_{N,t}^{(k)}\right]
\nonumber \\
&-i\frac{\d}{\d\alpha^*}\left[\left(\omega_p\alpha^*+\xi(t)
+\sum_r\lambda_r^*\beta_r^*\right)Q_{N,t}^{(k)}\right]
\nonumber \\
&+i\sum_r\left\{\frac{\d}{\d\beta_r}\left[\left(\omega_r\beta_r+\lambda_r\alpha\right)Q_{N,t}^{(k)}\right]
-\frac{\d}{\d\beta_r^*}\left[\left(\omega_r\beta_r^*+\lambda_r^*\alpha^*\right)Q_{N,t}^{(k)}\right]\right\},
\end{align}
which corresponds to the free time evolution of the 0-subsystem, and
\beq
-i{\cal L}_{\rm a}^{(k)}(\alpha)Q_{N,t}^{(k)}
:=-i\left[\sum_{j=1}^k\left(\omega_aS_j^z+H_{{\rm B}_j}+H_{{\rm I}_j}\right)
+g\sum_{j=1}^k\left(\alpha X_j^{\dagger}+\alpha^*X_j\right),Q_{N,t}^{(k)}\right],
\eeq
which represents the time evolution for $k$ atoms and environments 
under the ``effective field'' $g\sum_{j=1}^k\left(\alpha X_j^{\dagger}+\alpha^*X_j\right)$.
The super-operators ${\cal V}^{(k)}$ and ${\cal W}^{(k)}$ are defined by
\begin{align}
{\cal V}^{(k)}Q_{N,t}^{(k)}&:=ig\frac{\d}{\d\alpha}Q_{N,t}^{(k)}\sum_{j=1}^kX_j-ig\sum_{j=1}^kX_j^{\dagger}\frac{\d}{\d\alpha^*}Q_{N,t}^{(k)},
\label{eq:V}
\\
{\cal W}^{(k)}Q_{N,t}^{(k+1)}&:=ig\frac{\d}{\d\alpha}{\rm Tr}_{k+1}X_{k+1}Q_{N,t}^{(k+1)}
-ig\frac{\d}{\d\alpha^*}{\rm Tr}_{k+1}X_{k+1}^{\dagger}Q_{N,t}^{(k+1)}.
\label{eq:W}
\end{align}

As was mentioned in Sec.~\ref{sec:model}, it is remarked that if we consider the case in which the 0-environment acts as a thermal bath,
the limit of $\Lambda\rightarrow\infty$ should be taken {\it before} $N\rightarrow\infty$.
By assumption, $\{\lambda_r\}$ satisfies $\sum_r|\lambda_r|^2<\infty$ in the limit of $\Lambda\rightarrow\infty$.
In this case, the state of the $0$-environment is almost unchanged, that is, $\< b_r\>\sim \rt{N}\beta_r\sim O(\rt{N}\lambda_r)\sim O(\rt{N/\Lambda})$, 
if the $0$-environment is initially in equilibrium, so it is consistent with the interpretation that the 0-environment is a thermal reservoir.
On the other hand, as is seen in the above estimation, when $\Lambda\sim O(N)$, the state of the 0-environment is strongly disturbed,
and it cannot be regarded as a thermal bath in this case.
Anyway, the justification of the MF approach is possible for both these two cases.

If we formally take the limit of $N\rightarrow\infty$ in Eq.~(\ref{eq:dQdt}), we obtain the following equation of motion:
\beq
\frac{d}{dt}Q_{{\rm MF},t}^{(k)}=-i({\cal L}_{\rm p}(t)+{\cal L}_{\rm a}^{(k)}(\alpha))Q_{{\rm MF},t}^{(k)}
+{\cal W}^{(k)}Q_{{\rm MF},t}^{(k+1)}.
\label{eq:dQdtMF}
\eeq
If initially the density matrix is in the product form,
$$Q_{{\rm MF},0}^{(k)}=Q_{{\rm MF},0}^{(0)}\otimes(\rho_{\rm a})_0^{\otimes k}, \quad \forall k\in {\mathbb N},$$
the solution of Eq.~(\ref{eq:dQdtMF}) is equivalent to that of the MF dynamical equation.
Therefore, we shall compare the solutions of Eq.~(\ref{eq:dQdt}) with those of Eq.~(\ref{eq:dQdtMF}).

We define the time evolution operator due to $-i{\cal L}_{\rm p}(t)-i{\cal L}_{\rm a}^{(k)}(\alpha)$ as
\beq
U_{t,s}^{(k)}:=\overleftarrow{\rm T}\exp\left[-i\int_s^tdt'\left({\cal L}_{\rm p}(t')+{\cal L}_{\rm a}^{(k)}(\alpha)\right)\right].
\eeq
Here $\overleftarrow{\rm T}$ is the time-ordering operator (the arrow implies the direction from past to future).
Then we can obtain the formal integral equation for $Q_{N,t}^{(k)}$ from Eq.~(\ref{eq:dQdt}) as
\begin{align}
Q_{N,t}^{(k)}=&U_{t,0}^{(k)}Q_{N,0}^{(k)}(0)
+\int_0^tdt_1U_{t,t_1}^{(k)}{\cal W}^{(k)}Q_{N,t_1}^{(k+1)}
\nonumber \\
&+\frac{1}{N}\int_0^tdt_1U_{t,t_1}^{(k)}{\cal V}^{(k)}Q_{N,t_1}^{(k)}
-\frac{k}{N}\int_0^tdt_1U_{t,t_1}^{(k)}{\cal W}^{(k)}Q_{N,t_1}^{(k+1)}.
\end{align}
It is almost the same as Eq.~(\ref{eq:d_BBGKY}).
By substituting this expression into the RHS iteratively $L$ times,
we obtain the following Duhamel series expansion:
\begin{align}
&Q_{N,t}^{(k)}=\sum_{l=0}^{L-1}\int_0^tdt_1\int_0^{t_1}dt_2\dots\int_0^{t_{l-1}}dt_l
U_{t,t_1}^{(k)}{\cal W}^{(k)}U_{t_1,t_2}^{(k+1)}{\cal W}^{(k+1)}
\dots {\cal W}^{k+l-1}U_{t_l,0}^{(k+l)}Q_{N,0}^{(k+l)}
\nonumber \\
&+\int_0^tdt_1\int_0^{t_1}dt_2\dots\int_0^{t_{L-1}}dt_LU_{t,t_1}^{(k)}{\cal W}^{(k)}U_{t_1,t_2}^{(k+1)}{\cal W}^{(k+1)}
\dots U_{t_{L-1},t_L}^{(k+L-1)}{\cal W}^{(k+L-1)}Q_{N,t_L}^{(k+L)}
\nonumber \\
&+\frac{1}{N}\sum_{l=1}^L\int_0^tdt_1\int_0^{t_1}dt_2\dots\int_0^{t_{l-1}}dt_l
U_{t,t_1}^{(k)}{\cal W}^{(k)}U_{t_1,t_2}^{(k+1)}{\cal W}^{(k+1)}\dots U_{t_{l-1},t_l}^{(k+l-1)}{\cal V}^{(k+l-1)}Q_{N,t_l}^{(k+l-1)}
\nonumber \\
&-\sum_{l=1}^L\frac{k+l-1}{N}\int_0^tdt_1\int_0^{t_1}dt_2\dots\int_0^{t_{l-1}}dt_l
U_{t,t_1}^{(k)}{\cal W}^{(k)}U_{t_1,t_2}^{(k+1)}{\cal W}^{(k+1)}
\nonumber \\
&\qquad\qquad\qquad\qquad\qquad\qquad\qquad\qquad\qquad\dots U_{t_{l-1},t_l}^{(k+l-1)}{\cal W}^{(k+l-1)}Q_{N,t_l}^{(k+l)}.
\label{eq:BBGKY}
\end{align}

Similarly, we can obtain the Duhamel expansion for the MF density matrix $Q_{{\rm MF},t}^{(k)}$:
\begin{align}
Q_{{\rm MF},t}^{(k)}=\sum_{l=0}^{L-1}\int_0^tdt_1\int_0^{t_1}dt_2\dots\int_0^{t_{l-1}}dt_l
U_{t,t_1}^{(k)}{\cal W}^{(k)}U_{t_1,t_2}^{(k+1)}{\cal W}^{(k+1)}
\dots {\cal W}^{k+l-1}U_{t_l,0}^{(k+l)}Q_{{\rm MF},0}^{(k+l)}
\nonumber \\
+\int_0^tdt_1\int_0^{t_1}dt_2\dots\int_0^{t_{L-1}}dt_LU_{t,t_1}^{(k)}{\cal W}^{(k)}U_{t_1,t_2}^{(k+1)}{\cal W}^{(k+1)}
\dots U_{t_{L-1},t_L}^{(k+L-1)}{\cal W}^{(k+L-1)}Q_{{\rm MF},t_L}^{(k+L)}.
\label{eq:BBGKY_MF}
\end{align}

These expansions are the starting point for justification of the MF theory.

\subsection{Restriction of observables}
\label{sec:observable}

Let ${\cal O}^{(k)}$ be an operator acting on the Hilbert space $\bigotimes_{j=0}^k{\cal H}_j$.
Because the creation and annihilation operators for bosons are unbounded operators, it is hard to prove that
the expectation value of an {\it arbitrary} operator ${\cal O}^{(k)}$ 
calculated by $Q_{N,t}^{(k)}$ coincides with that calculated by $Q_{{\rm MF},t}$.
Therefore, we now restrict the physical quantities.

In this paper, we focus only on the observables ${\cal O}^{(k)}(\alpha,\alpha^*,\{\beta_r,\beta_r^*\})\in {\cal B}$.
The set of observables ${\cal B}$ is defined as follows.
All the observables ${\cal O}^{(k)}(\alpha,\alpha^*,\{\beta_r,\beta_r^*\})\in {\cal B}$ satisfy the following conditions.
There exist some complex numbers $c_0, c_a, \{ c_r\}\in\mathbb{C}$, with $\lim_{\Lambda\rightarrow\infty}\sum_r|c_r|^2<+\infty$,
and some positive integer $s>0$ and some positive numbers $\kappa>0$ and $\{ a_q>0\}$, $q=0,1,\dots, s$, such that
\beq
\left\|\frac{\d^m{\cal O}^{(k)}(\alpha,\alpha^*,\{\beta_r,\beta_r^*\})}
{\d\alpha_{\sigma_1}\d\alpha_{\sigma_2}\dots\d\alpha_{\sigma_m}}\right\|_{\infty}
\leq \kappa^m\sum_{q=0}^sa_q\left|c_0+c_a\alpha+\sum_rc_r\beta_r\right|^q,
\label{eq:restriction}
\eeq
for all $m=0,1,\dots$ and for all $\{\sigma_j\}_{j=1}^m$.
Here $\sigma_j=\pm 1$, $\alpha_1:=\alpha$, and $\alpha_{-1}:=\alpha^*$. 
For instance, all the operators expressed by a polynomial of $\left\{a, a^{\dagger}, \left\{\vec{S}_i\right\}_{i=1}^k\right\}$ belong to ${\cal B}$
after the correspondence $\{a/\rt{N}\rightarrow\alpha, a^{\dagger}/\rt{N}\rightarrow\alpha^*\}$ is made.

The derivatives $\d/\d\alpha$ and $\d/\d\alpha^*$ appear because of the quantum fluctuation, i.e. the commutation relation $[a,a^{\dagger}]=1$.
Roughly speaking, the condition~(\ref{eq:restriction}) means that we exclude observables that are too sensitive to quantum fluctuations.
For example, an operator like $\exp[\epsilon a^{\dagger}a/N]$, which roughly corresponds to ${\cal O}^{(k)}=\exp[\epsilon\alpha^*\alpha]$,
is excluded for any value of $\epsilon>0$ although its expectation value might exist and be finite.

\subsection{Free time evolution of the 0-subsystem}

In this section, we introduce several time evolution operators and discuss their relation.
The relation given by Eq.~(\ref{eq:U_property}) corresponds to the transformation from the ``Schr\"odinger picture'' to the ``interaction picture''
in terms of the coherent state representation.

Let us define
\beq
U_{t,s}^{\rm (p)}:=\overleftarrow{\rm T}\exp\left[-i\int_s^tdt'{\cal L}_{\rm p}(t')\right].
\eeq
For an arbitrary function $f(\alpha,\alpha^*,\{\beta_r,\beta_r^*\})$,
\beq
f(\alpha,\alpha^*,\{\beta_r,\beta_r^*\})U_{t,0}^{\rm (p)}
=U_{t,0}^{\rm (p)}f(\alpha(t),\alpha^*(t),\{\beta_r(t),\beta_r^*(t)\}),
\label{eq:Up}
\eeq
where $\alpha(t)$ is the solution of the following equations:
\beq
\left\{
\begin{aligned}
\frac{d}{dt}\alpha(t)&=-i\left[\omega_{\rm p}\alpha(t)+\xi(t)+\sum_r\lambda_r\beta_r(t)\right], \\
\frac{d}{dt}\beta_r(t)&=-i\left(\omega_r\beta_r(t)+\lambda_r\alpha(t)\right),
\end{aligned}
\right.
\label{eq:alpha_beta}
\eeq
with the initial condition $\alpha(0)=\alpha$ and $\beta_r(0)=\beta_r$.

The following property derived from Eq.~(\ref{eq:Up}) is important:
\beq
U_{t,s}^{(k)}U_{s,0}^{\rm (p)}=U_{t,0}^{\rm (p)}\tilde{U}_{t,s}^{(k)},
\label{eq:U_property}
\eeq
where
\beq
\tilde{U}_{t,s}^{(k)}:=\overleftarrow{T}\exp\left[-i\int_s^tdt'{\cal L}_{\rm a}^{(k)}(\alpha(t'))\right].
\eeq
This relation allows us to divide the time evolution operator $U_{t,s}^{(k)}$ into two parts, 
the free time evolution of the 0-subsystem $U_{t,0}^{\rm (p)}$ and the time evolution of the remaining part $\tilde{U}_{t,s}^{(k)}$.
This is viewed as the transformation to the ``interaction picture'' 
in the sense that the photon amplitude $\alpha(t)$ appearing in $k$ atoms' time evolution operator $\tilde{U}_{t,s}^{(k)}$
evolves under the Hamiltonian of the 0-subsystem.

Because these equations are linear, the solutions are written in the form
\beq
\left\{
\begin{aligned}
\alpha(t)=h_a(t)+g_{aa}(t)\alpha+\sum_rg_{ar}(t)\beta_r, \\
\beta_r(t)=h_r(t)+g_{ra}(t)\alpha+\sum_{r'}g_{rr'}(t)\beta_{r'}.
\end{aligned}
\right.
\label{eq:solution}
\eeq
It is noted that
\beq
\left\{
\begin{aligned}
|g_{aa}(t)|^2+\sum_r\left|g_{ar}(t)\right|^2=1, \\
|g_{ra}(t)|^2+\sum_{r'}\left|g_{rr'}(t)\right|^2=1.
\end{aligned}
\right.
\label{eq:bound_g}
\eeq
The matrix $g(t)$ is a unitary and symmetric matrix.
This property will be used later.
Here it should be noted that it is an important assumption that the 0-subsystem is a linear Bose system.

\section{Exactness of the MF dynamics in the optical cavity system}
\label{sec:statement}

From now on, we prove that the MF theory is exact in the limit of $N\rightarrow\infty$ in the sense that
if initially 
\beq
\lim_{N\rightarrow\infty}\int d^2\alpha\left(\prod_r\int d^2\beta_r\right){\rm Tr}_{1,2,\dots,k}
{\cal O}^{(k)}\left(\alpha,\alpha^*,\{\beta_r,\beta_r^*\}\right)\left(Q_{N,0}^{(k)}-Q_{{\rm MF},0}^{(k)}\right)=0
\label{eq:initial}
\eeq
for $\forall {\cal O}^{(k)}\in{\cal B}$ and any fixed $k\in\mathbb{N}$,
it implies that
\beq
\lim_{N\rightarrow\infty}\int d^2\alpha\left(\prod_r\int d^2\beta_r\right){\rm Tr}_{1,2,\dots,k}
{\cal O}^{(k)}\left(\alpha,\alpha^*,\{\beta_r,\beta_r^*\}\right)\left(Q_{N,t}^{(k)}-Q_{{\rm MF},t}^{(k)}\right)=0
\label{eq:proposition}
\eeq
for any fixed time $t>0$.

\subsection{Finiteness of expectation values of observables}
\label{sec:finite}

First of all, we prove that all the expectation values of observables ${\cal O}^{(k)}\in{\cal B}$ at time $t$ are finite
as long as they are also finite at initial time $t=0$.
Unboundedness of operators is obstructive for our proof, hence this property is desirable.
Once we can show this property, the justification of the MF approximation is almost straightforward.

From Eq.~(\ref{eq:BBGKY}), we obtain
\beq
\left|\int d^2\alpha\left(\prod_r\int d^2\beta_r\right){\rm Tr}_{1,2,\dots,k}
{\cal O}^{(k)}\left(\alpha,\alpha^*,\{\beta_r,\beta_r^*\}\right)Q_{N,t}^{(k)}\right|\leq B_1+B_2+B_3+B_4.
\label{eq:bound}
\eeq
$\{B_i\}$ are defined by
\begin{align*}
&B_1=\sum_{l=0}^{L-1}\left|\int d^2\alpha\left(\prod_r\int d^2\beta_r\right){\rm Tr}_{1,2,\dots,k}
{\cal O}^{(k)}\left(\alpha,\alpha^*,\{\beta_r,\beta_r^*\}\right)\right.
\\
&\left.\times\int_0^tdt_1\int_0^{t_1}dt_2\dots\int_0^{t_{l-1}}dt_l
U_{t,t_1}^{(k)}{\cal W}^{(k)}U_{t_1,t_2}^{(k+1)}{\cal W}^{(k+1)}
\dots {\cal W}^{(k+l-1)}U_{t_l,0}^{(k+l)}Q_{N,0}^{(k+l)}\right|,
\\
&B_2=\left|\int d^2\alpha\left(\prod_r\int d^2\beta_r\right){\rm Tr}_{1,2,\dots,k}
{\cal O}^{(k)}\left(\alpha,\alpha^*,\{\beta_r,\beta_r^*\}\right)\right.
\\
&\left.\times\int_0^tdt_1\int_0^{t_1}dt_2\dots\int_0^{t_{L-1}}dt_LU_{t,t_1}^{(k)}{\cal W}^{(k)}U_{t_1,t_2}^{(k+1)}{\cal W}^{(k+1)}
\dots U_{t_{L-1},t_L}^{(k+L-1)}{\cal W}^{(k+L-1)}Q_{N,t_L}^{(k+L)}\right|,
\end{align*}
\begin{align*}
&B_3=\frac{1}{N}\sum_{l=1}^L\left|\int d^2\alpha\left(\prod_r\int d^2\beta_r\right){\rm Tr}_{1,2,\dots,k}
{\cal O}^{(k)}\left(\alpha,\alpha^*,\{\beta_r,\beta_r^*\}\right)\right.
\\
&\left.\times\int_0^tdt_1\int_0^{t_1}dt_2\dots\int_0^{t_{l-1}}dt_l
U_{t,t_1}^{(k)}{\cal W}^{(k)}U_{t_1,t_2}^{(k+1)}{\cal W}^{(k+1)}\dots U_{t_{l-1},t_l}^{(k+l-1)}{\cal V}^{(k+l-1)}Q_{N,t_l}^{(k+l-1)}\right|,
\\
&B_4=\sum_{l=1}^L\frac{k+l-1}{N}\left|\int d^2\alpha\left(\prod_r\int d^2\beta_r\right){\rm Tr}_{1,2,\dots,k}
{\cal O}^{(k)}\left(\alpha,\alpha^*,\{\beta_r,\beta_r^*\}\right)\right.
\\
&\left.\times\int_0^tdt_1\int_0^{t_1}dt_2\dots\int_0^{t_{l-1}}dt_l
U_{t,t_1}^{(k)}{\cal W}^{(k)}U_{t_1,t_2}^{(k+1)}{\cal W}^{(k+1)}\dots U_{t_{l-1},t_l}^{(k+l-1)}{\cal W}^{(k+l-1)}Q_{N,t_l}^{(k+l)}\right|.
\end{align*}

What we have to do is to evaluate the upper bounds of $B_i$ ($i=1,\dots,4$).
Since the derivation is very complicated, we give the derivation in Appendix~\ref{appendix:bound_B},
and here we just mention the strategy briefly.
The analysis is similar to that in Sec.~\ref{sec:spin}, but due to the derivatives with respect to $\alpha$ and $\alpha^*$
in ${\cal V}^{(k)}$, Eq.~(\ref{eq:V}), and ${\cal W}^{(k)}$, Eq.~(\ref{eq:W}), it must be modified.
First, we move to the ``interaction picture'' by using Eq.~(\ref{eq:U_property}).
By integrating by part repeatedly, we rewrite $\{ B_i\}$ 
so that all the derivatives are acting only on $\{\tilde{U}_{t_n,t_{n+1}}^{(k+n)}\}$ ($n=1,2,\dots,l$ or $L$) and ${\cal O}^{(k)}$.
Then we can show that the derivatives of $\tilde{U}_{t_n,t_{n+1}}^{(k+n)}$ with respect to $\alpha$ and $\alpha^*$ are bounded above by Eq.~(\ref{eq:U_diff}).
In addition, the derivatives of ${\cal O}^{(k)}$ are also bounded due to the restriction of observables, see Eq.~(\ref{eq:restriction}).
By using these bounds, we can obtain the upper bounds of $\{ B_i\}$ which approach zero in the limit of $1\ll L\ll N$.

The derived upper bounds are the following:
\begin{align}
B_1&\leq 2^ke^{2\kappa}\int d^2\alpha\left(\prod_r\int d^2\beta_r\right)
\sum_{q=0}^sa_q\left|c_0+c_a\alpha(t)+\sum_rc_r\beta_r(t)\right|^qQ_{N,0}^{(0)}, \\
B_2&\leq 2^{-L+k-1}e^{2\kappa}
\sum_{q=0}^sa_q\max_{t'\in[0,t]}\int d^2\alpha\left(\prod_r\int d^2\beta_r\right)\left|c_0+c_a\alpha(t)+\sum_rc_r\beta_r(t)\right|^qQ_{N,t'}^{(0)}, \\
B_3&\leq\frac{1}{N}2^{k-1}(k+2)e^{2\kappa}\sum_{q=0}^sa_q
\max_{t'\in[0,t]}\int d^2\alpha\left(\prod_r\int d^2\beta_r\right)\left|c_0+c_a\alpha(t)+\sum_rc_r\beta_r(t)\right|^qQ_{N,t'}^{(0)},\\
B_4&\leq\frac{1}{N}2^{k-1}(k+1)e^{2\kappa}\sum_{q=0}^sa_q
\max_{t'\in[0,t]}\int d^2\alpha\left(\prod_r\int d^2\beta_r\right)\left|c_0+c_a\alpha(t)+\sum_rc_r\beta_r(t)\right|^qQ_{N,t'}^{(0)}.
\end{align}

From the derived upper bounds of $B_i$ ($i=1-4$),
we show that the expectation value of any observable in ${\cal B}$ at any fixed time $t$ is finite.
Collecting the derived upper bounds presented above, we obtain
\begin{align}
\left|\<{\cal O}^{(k)}\>_t\right|
<&2^ke^{2\kappa}\sum_{q=0}^sa_q\left\<\left|c_0+c_a\alpha(t)+\sum_rc_r\beta_r(t)\right|^q\right\>_0
\nonumber \\
&+2^{k-1}e^{2\kappa}\left(2^{-L}+\frac{2k+3}{N}\right)
\sum_{q=0}^sa_q\max_{t'\in[0,t]}\left\<\left|c_0+c_a\alpha(t)+\sum_rc_r\beta_r(t)\right|^q\right\>_{t'}
\nonumber \\
\equiv &2^ke^{2\kappa}\sum_{q=0}^sa_q\left\<\left|c_0+c_a\alpha(t)+\sum_rc_r\beta_r(t)\right|^q\right\>_0
\nonumber \\
&+\sum_{q=0}^sb_q\max_{t'\in[0,t]}\left\<\left|c_0+c_a\alpha(t)+\sum_rc_r\beta_r(t)\right|^q\right\>_{t'},
\label{eq:finite}
\end{align}
where $b_q=2^{k-1}e^{2\kappa}(2^{-L}+(2k+3)/N)a_q$.
We can see that $b_q\rightarrow 0$ in the limit of $N\rightarrow\infty$ and $L\rightarrow\infty$.
Therefore, in the limit of $N\rightarrow\infty$, we can choose $b_q$ as an arbitrarily small value.
Here $\<\cdot\>_t:=\int d^2\alpha(\prod\int d^2\beta_r)(\cdot)Q_{N,t}^{(0)}$ is shorthand notation of the expectation value.
In order to show the finiteness, we must show that
$$\max_{t'\in[0,t]}\left\<\left|c_0+c_a\alpha(t)+\sum_rc_r\beta_r(t)\right|^q\right\>_{t'}$$
is finite.

In order to do so, we put 
\beq
{\cal O}^{(k)}=\left|d_0+d_a\alpha+\sum_rd_r\beta_r\right|^n=:{\cal O}_n,
\eeq
and define
\beq
{\cal O}_n(t):=\left|d_0+d_a\alpha(t)+\sum_rd_r\beta_r(t)\right|^n.
\eeq
Because the constants $d_0$, $d_a$, and $\sum_r|d_r|^2$ are finite, the generality is not lost if we restrict these constants as
\beq
|\vec{d}|^2:=|d_0|^2+|d_a|^2+\sum_r|d_r|^2\leq 1.
\label{eq:c_restriction}
\eeq

Obviously, ${\cal O}_n\in{\cal B}$ with $s=n$, $c_0=d_0$, $c_a=d_a$, $c_r=d_r$, $\kappa=nd_a$, and $a_q=1$.

Now we evaluate the quantity $${\cal X}_n:=\max_{\vec{d}:|\vec{d}|\leq 1}\max_{t'\in[0,t]}\<{\cal O}_n\>_{t'}.$$
From Eq.~(\ref{eq:finite}), we have
\beq
{\cal X}_n<2^ke^{2\kappa}\sum_{q=0}^n\left\<{\cal O}_q(t)\right\>_0
+b\sum_{q=0}^n\max_{\vec{d}:|\vec{d}|\leq 1}\max_{t'\in[0,t]}\left\<{\cal O}_q(t)\right\>_{t'},
\label{eq:X_n}
\eeq
where $b=2^{k-1}e^{2\kappa}[2^{-L}+(2k+3)/N]$.

Now we use the property of $\alpha(t)$ and $\beta_r(t)$.
By substituting Eq.~(\ref{eq:solution}) into the definition of ${\cal O}_q(t)$, we have
\beq
{\cal O}_q(t)=\left|d_0'+d_a'\alpha+\sum_rd_r'\beta_r\right|^q,
\label{eq:O_q}
\eeq
where
\begin{equation*}
\left\{
\begin{aligned}
d_0'&=d_0+d_ah_a(t)+\sum_rd_rh_r(t)=:\vec{d}\cdot\vec{h}(t), \\
d_a'&=d_ag_{aa}(t)+\sum_rd_rg_{ra}(t), \\
d_r'&=d_ag_{ar}(t)+\sum_{r'}d_{r'}g_{r'r}(t).
\end{aligned}
\right.
\end{equation*}
The norm of this new vector $\vec{d'}$ is given by
$$|\vec{d'}|^2=|d_0'|^2+|d_a'|^2+\sum_r|d_r'|^2
\leq |\vec{h}(t)|^2+\sum_x\left|\sum_{x'}d_{x'}g_{x'x}(t)\right|^2,$$
where $x, x'\in\left\{a,\{r\}\right\}$.
Now we use the fact that the matrix $g(t)$ is unitary (the absolute value of any eigenvalue of $g(t)$ is unity).
This yields
\beq
|\vec{d'}|^2\leq |\vec{h}(t)|^2+1,
\eeq
and the RHS is finite for any fixed time $t$ because of the linearity of Eq.~(\ref{eq:alpha_beta}) and the assumption of non-pathological external field.
This also indicates that ${\cal O}_q(t)\in{\cal B}$, therefore, the first term of the RHS in Eq.~(\ref{eq:X_n}) is finite.
We write it as $f_n(t)=2^ke^{2\kappa}\sum_{q=0}^n\left\<{\cal O}_q(t)\right\>_0$.

We define the vector $\vec{d''}:=\vec{d'}/(|\vec{h}(t)|^2+1)$, whose norm is less than or equal to unity, $|\vec{d''}|\leq 1$
regardless of the choice of $\vec{d}$.
Obviously $\left\{\vec{d''}:|\vec{d''}|\leq 1\right\}\supseteq\left\{\vec{d''}:|\vec{d}|\leq 1\right\}$.
Therefore, from Eqs.~(\ref{eq:X_n}) and (\ref{eq:O_q}), we obtain
\begin{align}
{\cal X}_n&<f_n(t)+b\left(|\vec{h}(t)|^2+1\right)\sum_{q=0}^n\max_{\vec{d}:|\vec{d}|\leq 1}\max_{t'\in[0,t]}
\left\<\left|d_0''+d_a''\alpha+\sum_rd_r''\beta_r\right|^q\right\>_{t'}
\nonumber \\
&\leq f_n(t)+b\left(|\vec{h}(t)|^2+1\right)\sum_{q=0}^n\max_{\vec{d''}:|\vec{d''}|\leq 1}\max_{t'\in[0,t]}
\left\<\left|d_0''+d_a''\alpha+\sum_rd_r''\beta_r\right|^q\right\>_{t'}
\nonumber \\
&=f_n(t)+b\left(|\vec{h}(t)|^2+1\right)\sum_{q=0}^n{\cal X}_q
\end{align}
Since we can choose $b$ as an arbitrarily small number, by starting from ${\cal X}_0=1$,
we can inductively show that all the ${\cal X}_n$ are finite by using the above inequality.
This completes the proof of the finiteness of the expectation values, because the RHS of Eq.~(\ref{eq:finite}) is then finite.

In particular, we find
\beq
\lim_{L\rightarrow\infty}\lim_{N\rightarrow\infty}B_2=
\lim_{L\rightarrow\infty}\lim_{N\rightarrow\infty}B_3=
\lim_{L\rightarrow\infty}\lim_{N\rightarrow\infty}B_4=0.
\eeq

\subsection{Justification of the mean-field approach}
\label{sec:justification}

The proof proceeds very similarly to Sec~\ref{sec:finite}.
From Eqs.~(\ref{eq:BBGKY}) and (\ref{eq:BBGKY_MF}), we obtain the upper bound
\begin{align}
\left|\int d^2\alpha\left(\prod_r\int d^2\beta_r\right){\rm Tr}_{1,2,\dots,k}
{\cal O}^{(k)}\left(\alpha,\alpha^*,\{\beta_r,\beta_r^*\}\right)\left(Q_{N,t}^{(k)}-Q_{{\rm MF},t}^{(k)}\right)\right|
\nonumber \\
\leq B_1'+B_2'+B_3+B_4.
\end{align}
Here $B_3$ and $B_4$ are the same as in Sec.~\ref{sec:finite}. 
$B_1'$ and $B_2'$ are defined by
\begin{align}
&B_1':=\sum_{l=0}^{L-1}\left|\int d^2\alpha\left(\prod_r\int d^2\beta_r\right){\rm Tr}_{1,2,\dots,k}
{\cal O}^{(k)}\left(\alpha,\alpha^*,\{\beta_r,\beta_r^*\}\right)\right.
\nonumber \\
&\left.\times\int_0^tdt_1\int_0^{t_1}dt_2\dots\int_0^{t_{l-1}}dt_l
U_{t,t_1}^{(k)}{\cal W}^{(k)}U_{t_1,t_2}^{(k+1)}{\cal W}^{(k+1)}
\dots {\cal W}^{k+l-1}U_{t_l,0}^{(k+l)}\left(Q_{N,0}^{(k+l)}-Q_{{\rm MF},0}^{(k+l)}\right)\right|
\nonumber \\
&B_2':=\left|\int d^2\alpha\left(\prod_r\int d^2\beta_r\right){\rm Tr}_{1,2,\dots,k}
{\cal O}^{(k)}\left(\alpha,\alpha^*,\{\beta_r,\beta_r^*\}\right)\right.
\nonumber \\
&\times\int_0^tdt_1\int_0^{t_1}dt_2\dots\int_0^{t_{L-1}}dt_LU_{t,t_1}^{(k)}{\cal W}^{(k)}U_{t_1,t_2}^{(k+1)}{\cal W}^{(k+1)}
\nonumber \\
&\qquad\qquad\qquad\qquad\qquad\qquad
\dots U_{t_{L-1},t_L}^{(k+L-1)}{\cal W}^{(k+L-1)}\left(Q_{N,t_L}^{(k+L)}-Q_{{\rm MF},t_L}^{(k+L)}\right)\bigg|.
\nonumber \\
\end{align}

In Appendix~\ref{appendix:bound_B}, it is shown that $\lim_{N\rightarrow\infty}B_3=\lim_{N\rightarrow\infty}B_4=0$ in the time interval $t\in[0,t_0]$.
The proof of
\beq
\lim_{L\rightarrow\infty}\lim_{N\rightarrow\infty}B_1'=\lim_{L\rightarrow\infty}\lim_{N\rightarrow\infty}B_2'=0
\label{eq:B_prime}
\eeq
for $t\in[0,t_0]$ is similar to the analysis in Appendix~\ref{appendix:bound_B}.
We give the derivation of Eq.~(\ref{eq:B_prime}) in Appendix~\ref{appendix:bound_B_prime}.

Up to now, we showed that $B_1'+B_2'+B_3+B_4\rightarrow 0$ as $N\rightarrow\infty$ for $0\leq t\leq t_0$.
It means that Eq.~(\ref{eq:proposition}) has been shown for $0\leq t\leq t_0$.
Finally for completeness of the proof, we must extend the time interval from $t\in[0,t]$ to $t\in[0,\infty)$, but there is no difficulty on this point.
If we regard $t=t_0$ as a new initial time and repeat the argument, we can show that Eq.~(\ref{eq:proposition}) is correct for any fixed time $t>0$.
Thus the proof of the justification of the MF dynamics has been completed.

\section{Conclusion}
\label{sec:conclusion}

Finally we conclude the present work by making some remarks on our result:
\begin{itemize}
\item Our result is quite general: environments may be attached to the system of interest,
and a time-dependent driving field may be applied. 
When an environment is large enough, it acts as a thermal reservoir on the system of interest, hence the result of this work is applicable to open quantum systems.
However, we assumed a special Hamiltonian for the 0-environment, $H_{{\rm B}_0}+H_{{\rm I}_0}$ in Eq.~(\ref{eq:model}).
It is preferable to generalize our result to a wider class of 0-environmental Hamiltonians in the future.
\item We restricted the class of physical quantities to $\cal B$ in the present work. This restriction, however, might be just a technical assumption.
Extension to a more general class of operators is an important issue.
Another important issue is to treat the fluctuations.
For instance, $a^{\dagger}a/N=(aa^{\dagger}-1)/N=(a/\rt{N})(a^{\dagger}/\rt{N})-1/N$ belongs to $\cal B$,
but $(a^{\dagger}a-\<a^{\dagger}\>\<a\>)/\rt{N}$, whose expectation value converges to a finite value as $N\rightarrow\infty$, does not.
In the seminal work by Hepp and Lieb~\cite{Hepp-Lieb1973}, the authors showed that,
under the singular reservoir limit and the approximation of replacing the bosons by the ensemble of fermions, 
the equations of motion for the fluctuations
are equal to the linearized equation of the intensive observables, 
which is known as the Onsager's regression hypothesis~\cite{Onsager1931-1,Onsager1931-2,Kubo_text}.
We have not been able to confirm the validity of this hypothesis without those approximations yet.
\item Without the driving force, and if the thermal reservoirs have identical temperatures,
the MF dynamics usually predicts thermalization of the system of interest.
This means that the limit of $N\rightarrow\infty$ and the limit of $t\rightarrow\infty$ are interchangeable under such an equilibrium situation.
In other words, there is no quasi-stationary state in open systems under the equilibrium situation, 
which is in contrast to closed systems, see~\cite{Gupta-Mukamel2010} for this aspect.
However, it is less obvious whether or not the thermodynamic limit and the long-time limit are interchangeable under the nonequilibrium situation studied in this paper.
In this work, we always take the limit of $N\rightarrow\infty$ first and then the limit of $t\rightarrow\infty$.
In a real experimental setup, there can be a situation in which the thermodynamic limit ought to be taken after the long-time limit
since the number of atoms in the cavity is not large enough.
\item We can justify the MF theory, but it is a separate issue whether the MF dynamical equations, Eqs.~(\ref{eq:MF_p}) and (\ref{eq:MF_a}), can be solved.
Since the degrees of freedom of environmental systems remain, it is difficult to solve the MF dynamical equations exactly.
It would be very interesting if we could exactly solve the quantum dynamics of the whole system including thermal reservoirs with the help of the MF theory.
If this were done, we would be able to obtain some insight 
into the effects of non-Markovian dynamics~\cite{Breuer_text} 
and the accuracy of the perturbative quantum master equation at long times~\cite{Mori2008,Fleming2011}
{\it in an interacting many body system}.
When the system of interest is small, which is the standard setting of open quantum systems,
it is recognized that the non-Markovian effect is negligible in the van Hove limit~\cite{Davies_text}.
When the system of interest is large, however, it is not obvious whether the use of the Born-Markov quantum master equation is justified 
even for the small coupling between the system of interest and the thermal bath,
because two limits, the thermodynamic limit and the van Hove limit, are involved.
Therefore, it is important to obtain the exact time evolution in a macroscopic open system.
It is noted that Merkli and Berman~\cite{Merkli-Berman2012} recently obtained a rigorous result in this direction for a simple model 
which is purely dephasing and where the relaxation of energy does not take place.
In our setting, if we assume that each spin interacts with infinitely many reservoirs of the same structure,
the dynamics will be solved at least in a numerically exact manner,
because we can also apply the MF approximation to the coupling between the system of interest and the reservoirs in that case.
\end{itemize}

We hope that the result of this work will become a good starting point to consider the above remaining problems
and analyze some interesting nonequilibrium phenomena.

\section*{Acknowledgments}
The author thanks Seiji Miyashita and Tatsuhiko Shirai for fruitful discussion.
This work is supported by the Sumitomo Foundation (grant No. 120753).

\appendix
\section{Upper bounds of $\{B_i\}$}
\label{appendix:bound_B}
\subsection*{Evaluation of $B_1$}

First we consider the term $B_1$.
From Eqs.~(\ref{eq:V}), (\ref{eq:W}), and (\ref{eq:U_property}),
we obtain the following expression:
\begin{align}
B_1=&\sum_{l=0}^{L-1}\left|\int d^2\alpha\left(\prod_r\int d^2\beta_r\right){\rm Tr}_{1,2,\dots,k}
{\cal O}^{(k)}(\alpha(t),\alpha^*(t))\int_0^tdt_1\int_0^{t_1}dt_2\dots\int_0^{t_{l-1}}dt_l\right.
\nonumber \\
&\quad\times\sum_{\{\sigma_j=\pm 1\}}^l
\tilde{U}_{t,t_1}^{(k)}\frac{\d}{\d\alpha_{\sigma_1}(t_1)}{\rm Tr}_{k+1}X_{k+1}^{(\sigma_1)}
\tilde{U}_{t_1,t_2}^{(k+1)}\frac{\d}{\d\alpha_{\sigma_2}(t_2)}{\rm Tr}_{k+2}X_{k+2}^{(\sigma_2)}
\nonumber \\
&\quad\left.\dots \frac{\d}{\d\alpha_{\sigma_l}(t_l)}{\rm Tr}_{k+l}X_{k+l}^{(\sigma_l)}
\tilde{U}_{t_l,0}^{(k+l)}Q_{N,0}^{(k+l)}\right|,
\end{align}
where $X_j^{(1)}:=X_j$ and $X_j^{(-1)}:=X_j^{\dagger}$.
By performing integration by part, we obtain
\begin{align}
B_1=&\sum_{l=0}^{L-1}\Bigg|\int_0^tdt_1\int_0^{t_1}dt_2\dots\int_0^{t_{l-1}}dt_l
\int d^2\alpha\left(\prod_r\int d^2\beta_r\right){\rm Tr}_{1,2,\dots,k+l}
\nonumber \\
&\quad\times\sum_{\{\sigma_j=\pm 1\}}^l X_{k+1}^{(\sigma_1)}X_{k+2}^{(\sigma_2)}\dots X_{k+l}^{(\sigma_l)}
\nonumber \\
&\quad\times
\frac{\d}{\d\alpha_{\sigma_l}(t_l)}\bigg\{\frac{\d}{\d\alpha_{\sigma_{l-1}}(t_{l-1})}\bigg\{
\dots \frac{\d}{\d\alpha_{\sigma_1}(t_1)}\bigg\{{\cal O}^{(k)}(\alpha(t),\alpha^*(t),\{\beta_r(t),\beta_r^*(t)\})
\nonumber \\
&\quad\times\tilde{U}_{t,t_1}^{(k)}\bigg\}
\tilde{U}_{t_1,t_2}^{(k+1)}\bigg\}\dots \tilde{U}_{t_{l-1},t_l}^{(k+l-1)}\bigg\}\tilde{U}_{t_l,0}^{(k+l)}
Q_{N,0}^{(k+l)}\Bigg|.
\end{align}
We put
\begin{align}
B_1^{(k+l)}:=&\int_0^tdt_1\int_0^{t_1}dt_2\dots\int_0^{t_{l-1}}dt_l
\sum_{\{\sigma_j=\pm 1\}}^l X_{k+1}^{(\sigma_1)}X_{k+2}^{(\sigma_2)}\dots X_{k+l}^{(\sigma_l)}
\nonumber \\
&\quad\times
\frac{\d}{\d\alpha_{\sigma_l}(t_l)}\bigg\{\frac{\d}{\d\alpha_{\sigma_{l-1}}(t_{l-1})}\bigg\{
\dots \frac{\d}{\d\alpha_{\sigma_1}(t_1)}\bigg\{
\nonumber \\
&\quad\times{\cal O}^{(k)}\left(\alpha(t),\alpha^*(t),\{\beta_r(t),\beta_r^*(t)\}\right)\tilde{U}_{t,t_1}^{(k)}\bigg\}
\tilde{U}_{t_1,t_2}^{(k+1)}\bigg\}\dots \tilde{U}_{t_{l-1},t_l}^{(k+l-1)}\bigg\}\tilde{U}_{t_l,0}^{(k+l)}.
\label{eq:A1_k+l}
\end{align}
Then $B_1$ is expressed as
\beq
B_1=\sum_{l=0}^{L-1}\left|
\int d^2\alpha\left(\prod_r\int d^2\beta_r\right){\rm Tr}_{1,2,\dots,k+l}B_1^{(k+l)}Q_{N,0}^{(k+l)}\right|.
\label{eq:A1_expression}
\eeq

In Eq.~(\ref{eq:A1_k+l}), a quantity like
$$\frac{\d^n\tilde{U}^{(k)}_{t,s}}{\d\alpha_{\sigma_1}(t_1)\d\alpha_{\sigma_2}(t_2)\dots\d\alpha_{\sigma_n}(t_n)}$$
appears.
We find that
\begin{align}
\frac{\d\tilde{U}^{(k)}_{t,s}}{\d\alpha_{\sigma_1}(t_1)}
&=-i\int_s^tdt'\tilde{U}_{t,t'}^{(k)}\frac{\d{\cal L}_{\rm a}(\alpha(t'))}{\d\alpha_{\sigma_1}(t_1)}\tilde{U}_{t',s}^{(k)}
\nonumber \\
&=-ig\int_s^tdt'\tilde{U}_{t,t'}^{(k)}\frac{\d\alpha_{\sigma_1}(t')}{\d\alpha_{\sigma_1}(t_1)}\sum_{j=1}^kX_j^{(-\sigma_1)\times}\tilde{U}_{t',s}^{(k)}
\nonumber \\
&=-ig\int_s^tdt'\tilde{U}_{t,t'}^{(k)}g_{aa}(t'-t_1)\sum_{j=1}^kX_j^{(-\sigma_1)\times}\tilde{U}_{t',s}^{(k)},
\end{align}
where $X_j^{(-\sigma_1)\times}(\cdot):=[X_j^{(-\sigma_1)},(\cdot)]$ is the commutator.
Remember that $|g_{aa}(t)|\leq 1$ because of Eq.~(\ref{eq:bound_g}).
Therefore,
\beq
\left\|\frac{\d\tilde{U}^{(k)}_{t,s}}{\d\alpha_{\sigma_1}(t_1)}\right\|_{\infty}
\leq 2kg\|X\|_{\infty}(t-s).
\eeq
Similarly, we can obtain
\beq
\left\|\frac{\d^n\tilde{U}^{(k)}_{t,s}}{\d\alpha_{\sigma_1}(t_1)\d\alpha_{\sigma_2}(t_2)\dots\d\alpha_{\sigma_n}(t_n)}\right\|_{\infty}
\leq\left[2kg\|X\|_{\infty}(t-s)\right]^n.
\label{eq:U_diff}
\eeq
Because of Eq.~(\ref{eq:U_diff}), 
we can replace $\d\tilde{U}_{t,s}/\d\alpha_{\sigma}(t')$ simply by $2kg\|X\|_{\infty}(t-s)\tilde{U}_{t,s}$ 
as far as it is concerned with the upper bound of the operator norm.
This fact makes much easier to evaluate the upper bound.

We define the set $${\cal N}_{l,n}:=\left\{\{m_1,m_2,\dots,m_n\}:1\leq m_1<m_2<\dots<m_n\leq l, \quad m_j\in{\mathbb N}\right\}.$$
If one of the elements of ${\cal N}_{l,n}$ is denoted by $N_{l,n}=\{m_1,m_2,\dots, m_n\}$,
we define $N^c_{l,n}:=\{1,2,\dots,l\}\backslash N_{l,n}$.

By using Eq.~(\ref{eq:U_diff}), we then obtain
\begin{align}
\left\|B_1^{(k+l)}\right\|_{\infty}
\leq&\frac{(g\|X\|_{\infty}t)^l}{l!}\sum_{\{\sigma_j=\pm 1\}}\sum_{n=0}^l\sum_{N_{l,n}\in{\cal N}_{l,n}}
\nonumber \\
&\times\left\|\left(\prod_{j\in N_{l,n}}\frac{\d}{\d\alpha_{\sigma_j}(t)}\right)
{\cal O}^{(k)}(\alpha(t),\alpha^*(t),\{\beta_r(t),\beta_r^*(t)\})\right\|_{\infty}
\nonumber \\
&\times (k+l-1)(k+l-2)\dots(k+n)\times(2g\|X\|_{\infty}t)^{l-n}.
\end{align}
Because ${\cal O}^{(k)}\in{\cal B}$, we have
\begin{align}
\left\|B_1^{(k+l)}\right\|_{\infty}
&\leq\frac{(g\|X\|_{\infty}t)^l}{l!}\sum_{\{\sigma_j=\pm 1\}}\sum_{n=0}^l\sum_{N_{l,n}\in{\cal N}_{l,n}}
\sum_{q=0}^sa_q\left|c_0+c_a\alpha(t)+\sum_rc_r\beta_r(t)\right|^q\kappa^n
\nonumber \\
&\times(k+l-1)(k+l-2)\dots(k+n)\times(2g\|X\|_{\infty}t)^{l-n}.
\end{align}
Since $\sum_{\{\sigma_j=\pm 1\}}1=2^l$, $\sum_{N_{l,n}\in{\cal N}_{l,n}}1=l!/(n!(l-n)!)$, we obtain the upper bound
\begin{align}
\left\|B_1^{(k+l)}\right\|_{\infty}
\leq& \sum_{n=0}^{\infty}\frac{\kappa^n}{n!}\frac{(k+l-1)!}{(l-n)!(k+n-1)!}(2g\| X\|_{\infty}t)^{2l-n}
\nonumber \\
&\times\sum_{q=0}^sa_q\left|c_0+c_a\alpha(t)+\sum_rc_r\beta_r(t)\right|^q.
\end{align}
Here we used $(k+l-1)!/[(l-n)!(k+n-1)!]\leq 2^{k+l-1}$ and 
restrict the time interval $t\in[0,t_0]$ where $t_0$ is determined by $2g\|X\|_{\infty}t_0=1/2$.
This restriction will be removed in Sec.~\ref{sec:justification}.
Then we obtain
\begin{align}
\left\|B_1^{(k+l)}\right\|_{\infty}
\leq 2^{k-l-1}e^{2\kappa}\sum_{q=0}^sa_q\left|c_0+c_a\alpha(t)+\sum_rc_r\beta_r(t)\right|^q.
\label{eq:A1_bound}
\end{align}

From Eq.~(\ref{eq:A1_expression}) and Eq.~(\ref{eq:A1_bound}), we obtain
\beq
B_1\leq 2^ke^{2\kappa}\int d^2\alpha\left(\prod_r\int d^2\beta_r\right)
\sum_{q=0}^sa_q\left|c_0+c_a\alpha(t)+\sum_rc_r\beta_r(t)\right|^qQ_{N,0}^{(0)}.
\eeq
Here, $\left|c_0+c_a\alpha(t)+\sum_rc_r\beta_r(t)\right|^q\in{\cal B}$ (which will be explicitly shown in Sec.~\ref{sec:finite}).
By assumption, at the initial time $t=0$, the expectation value is finite for any ${\cal O}^{(k)}\in{\cal B}$.
Therefore, $B_1$ is also finite.

\subsection*{Evaluation of $B_2$}

$B_2$ is given by
\begin{align}
B_2=\Bigg|\int d^2\alpha\left(\prod_r\int d^2\beta_r\right){\rm Tr}_{1,2,\dots,k}
{\cal O}^{(k)}\left(\alpha,\alpha^*,\{\beta_r,\beta_r^*\}\right)\int_0^tdt_1\int_0^{t_1}dt_2\dots\int_0^{t_{L-1}}dt_L
\nonumber \\
U_{t,t_1}^{(k)}{\cal W}^{(k)}U_{t_1,t_2}^{(k+1)}{\cal W}^{(k+1)}
\dots U_{t_{L-1},t_L}^{(k+L-1)}{\cal W}^{(k+L-1)}Q_{N,t_L}^{(k+L)}\Bigg|
\end{align}
This is written in the form
\beq
B_2\leq\int d^2\alpha\left(\prod_r\int d^2\beta_r\right)\int_0^tdt_1\int_0^{t_1}dt_2\dots\int_0^{t_{L-1}}dt_L
{\rm Tr}_{1,2,\dots,k+L}\left\|B_2^{(k+L)}\right\|_{\infty}Q_{N,t_L}^{(k+L)},
\label{eq:A2}
\eeq
where $B_2^{(k+L)}$ is given by
\begin{align}
B_2^{(k+L)}:=&
g^L\sum_{\{\sigma_j=\pm 1\}}^l X_{k+1}^{(\sigma_1)}X_{k+2}^{(\sigma_2)}\dots X_{k+L}^{(\sigma_L)}
\nonumber \\
&\quad\times
\frac{\d}{\d\alpha_{\sigma_L}(t_L)}\bigg\{\frac{\d}{\d\alpha_{\sigma_{L-1}}(t_{L-1})}\bigg\{
\dots \frac{\d}{\d\alpha_{\sigma_1}(t_1)}\bigg\{
\nonumber \\
&\quad\times{\cal O}^{(k)}\left(\alpha(t),\alpha^*(t),\{\beta_r(t),\beta_r^*(t)\}\right)\tilde{U}_{t,t_1}^{(k)}\bigg\}
\tilde{U}_{t_1,t_2}^{(k+1)}\bigg\}\dots \tilde{U}_{t_{L-1},t_L}^{(k+L-1)}\bigg\}.
\label{eq:A2_k+L}
\end{align}
Similarly to the evaluation of $B_1$, we obtain the upper bound
\begin{align}
B_2\leq&\int d^2\alpha\left(\prod_r\int d^2\beta_r\right)\int_0^tdt_1\dots\int_0^{t_L-1}dt_L
L!(2g\|X\|_{\infty})^L
\nonumber\\
&\times\sum_{m=0}^L\frac{1}{m!}2^{k+L-1}(2g\|X\|_{\infty}t)^{L-m}
\kappa^m\sum_{q=0}^sa_q\left|c_0+c_a\alpha(t)+\sum_rc_r\beta_r(t)\right|^qQ_{N,t_L}^{(k+L)}
\nonumber \\
\leq&(2g\|X\|_{\infty}t)^L\sum_{m=0}^L\frac{1}{m!}2^{k+L-1}(2g\|X\|_{\infty}t)^{L-m}
\nonumber\\
&\times\kappa^m\sum_{q=0}^sa_q
\max_{t'\in[0,t]}\int d^2\alpha\left(\prod_r\int d^2\beta_r\right)\left|c_0+c_a\alpha(t)+\sum_rc_r\beta_r(t)\right|^qQ_{N,t'}^{(k+L)}
\end{align}
Here, we again restrict the time interval $t\in[0,t_0]$ ($t_0$ was determined by $2g\|X\|_{\infty}t_0=1/2$ in the previous subsection).
Then we obtain
\begin{align}
\left\|B_2^{(k+L)}\right\|_{\infty}&\leq 2^{-L+k-1}\sum_{m=0}^L\frac{1}{m!}(2\kappa)^m
\sum_{q=0}a_q\left|c_0+c_a\alpha(t)+\sum_rc_r\beta_r(t)\right|^q
\nonumber \\
&\leq 2^{-L+k-1}e^{2\kappa}\sum_{q=0}^sa_q\left|c_0+c_a\alpha(t)+\sum_rc_r\beta_r(t)\right|^q
\label{eq:A2_k+L_bound}
\end{align}
Substituting it into Eq.~(\ref{eq:A2}), we have
\begin{align}
B_2\leq&2^{-L+k-1}\sum_{m=0}^L\frac{1}{m!}(2\kappa)^m
\sum_{q=0}^sa_q\max_{t'\in[0,t]}\int d^2\alpha\left(\prod_r\int d^2\beta_r\right)
\nonumber \\
&\qquad\qquad\qquad\qquad\qquad\qquad
\times\left|c_0+c_a\alpha(t)+\sum_rc_r\beta_r(t)\right|^qQ_{N,t'}^{(0)}
\nonumber \\
\leq& 2^{-L+k-1}e^{2\kappa}
\sum_{q=0}^sa_q\max_{t'\in[0,t]}\int d^2\alpha\left(\prod_r\int d^2\beta_r\right)\left|c_0+c_a\alpha(t)+\sum_rc_r\beta_r(t)\right|^qQ_{N,t'}^{(0)}
\label{eq:A2_bound}
\end{align}

\subsection*{Evaluation of $B_3$}

Next we evaluate $B_3$, which is given by
\begin{align}
B_3=&\frac{1}{N}\sum_{l=1}^L\left|\int d^2\alpha\left(\prod_r\int d^2\beta_r\right){\rm Tr}_{1,2,\dots,k}
{\cal O}^{(k)}\left(\alpha,\alpha^*,\{\beta_r,\beta_r^*\}\right)\right.
\nonumber \\
&\left.\times\int_0^tdt_1\int_0^{t_1}dt_2\dots\int_0^{t_{l-1}}dt_l
U_{t,t_1}^{(k)}{\cal W}^{(k)}U_{t_1,t_2}^{(k+1)}{\cal W}^{(k+1)}\dots U_{t_{l-1},t_l}^{(k+l-1)}{\cal V}^{(k+l-1)}Q_{N,t_l}^{(k+l-1)}\right|.
\end{align}
If we put
\begin{align}
B_3^{(k+l)}:=&(-ig)^lt\int_0^tdt_1\int_0^{t_1}dt_2\dots\int_0^{t_{l-1}}dt_{l}
\nonumber \\
&\times\sum_{\{\sigma_j=\pm 1\}} X_{k+1}^{(\sigma_1)}X_{k+2}^{(\sigma_2)}\dots X_{k+l}^{(\sigma_l)}
\nonumber \\
&\times
\frac{\d}{\d\alpha_{\sigma_l}(t_l)}\bigg\{\frac{\d}{\d\alpha_{\sigma_{l-1}}(t_{l-1})}\bigg\{
\dots \frac{\d}{\d\alpha_{\sigma_1}(t_1)}\bigg\{
\nonumber \\
&\times{\cal O}^{(k)}\left(\alpha(t),\alpha^*(t),\{\beta_r(t),\beta_r^*(t)\}\right)\tilde{U}_{t,t_1}^{(k)}\bigg\}
\tilde{U}_{t_1,t_2}^{(k+1)}\bigg\}\dots \tilde{U}_{t_{l-1},t_l}^{(k+l-1)}\bigg\}{\cal V}_{\sigma_l}^{(k+l)},
\end{align}
where
\beq
{\cal V}_{\sigma}^{(k)}(\cdot):=\left\{
\begin{aligned}
(\cdot)\sum_{j=1}^kX_j \quad &\text{for } \sigma=1, \\
\sum_{j=1}^kX_j^{\dagger}(\cdot) \quad &\text{for } \sigma=-1,
\end{aligned}
\right.
\eeq
we obtain the upper bound of $B_3$:
\beq
B_3\leq\frac{1}{N}\sum_{l=1}^L
\max_{t'\in[0,t]}\int d^2\alpha\left(\prod_r\int d^2\beta_r\right)\left\|B_3^{(k+l)}\right\|_{\infty}Q_{N,t'}^{(0)}.
\eeq
Similarly to the evaluation of $B_2$, we obtain for $0< t\leq t_0$
\beq
\left\|B_3^{(k+l)}\right\|_{\infty}\leq (k+l)2^{k-l-1}e^{2\kappa}\sum_{q=0}^sa_q\left|c_0+c_a\alpha(t)+\sum_rc_r\beta_r(t)\right|^q.
\eeq
Therefore,
\begin{align}
B_3\leq\frac{1}{N}\sum_{l=1}^L(k+l)2^{k-l-1}e^{2\kappa}
\sum_{q=0}^sa_q\max_{t'\in[0,t]}\int d^2\alpha\left(\prod_r\int d^2\beta_r\right)
\nonumber \\
\times\left|c_0+c_a\alpha(t)+\sum_rc_r\beta_r(t)\right|^qQ_{N,t_l}^{(0)}.
\end{align}
By elementary calculation, we find $\sum_{l=1}^Ll/2^l< 2$, and thus
\beq
B_3<\frac{1}{N}2^{k-1}(k+2)e^{2\kappa}\sum_{q=0}^sa_q
\max_{t'\in[0,t]}\int d^2\alpha\left(\prod_r\int d^2\beta_r\right)\left|c_0+c_a\alpha(t)+\sum_rc_r\beta_r(t)\right|^qQ_{N,t'}^{(0)}.
\label{eq:A3_bound}
\eeq

\subsection*{Evaluation of $B_4$}

Finally we evaluate the contribution of $B_4$. Its upper bound is given by
\begin{align}
B_4&\leq \frac{1}{N}\sum_{l=1}^L\max_{t'\in[0,t]}\int d^2\alpha\left(\prod_r\int d^2\beta_r\right)
\left\|B_4^{(k+l)}\right\|_{\infty}Q_{N,t'}^{(k+l)},
\end{align}
where $B_4^{(k+l)}$ is defined by
\begin{align}
B_4^{(k+l)}:=&(-ig)^l(k+l-1)
\int_0^tdt_1\int_0^{t_1}dt_2\dots\int_0^{t_{l-1}}dt_{l}
\nonumber \\
&\times\sum_{\{\sigma_j=\pm 1\}} X_{k+1}^{(\sigma_1)}X_{k+2}^{(\sigma_2)}\dots X_{k+l}^{(\sigma_l)}
\nonumber \\
&\times
\frac{\d}{\d\alpha_{\sigma_l}(t_l)}\bigg\{\frac{\d}{\d\alpha_{\sigma_{l-1}}(t_{l-1})}\bigg\{
\dots \frac{\d}{\d\alpha_{\sigma_1}(t_1)}\bigg\{
\nonumber \\
&\times{\cal O}^{(k)}\left(\alpha(t),\alpha^*(t),\{\beta_r(t),\beta_r^*(t)\}\right)\tilde{U}_{t,t_1}^{(k)}\bigg\}
\tilde{U}_{t_1,t_2}^{(k+1)}\bigg\}\dots \tilde{U}_{t_{l-1},t_l}^{(k+l-1)}\bigg\}.
\end{align}

We can evaluate the upper bound of $B_4^{(k+l)}$ by some calculations similar to those in previous subsections, and the result is
\beq
\left\|B_4^{(k+l)}\right\|_{\infty}
\leq (k+l-1)2^{k-l-1}e^{2\kappa}\sum_{q=0}^sa_q\left|c_0+c_a\alpha(t)+\sum_rc_r\beta_r(t)\right|^q,
\eeq
for $0\leq t\leq t_0$.
Therefore, we conclude that
\beq
B_4<\frac{1}{N}2^{k-1}(k+1)e^{2\kappa}\sum_{q=0}^sa_q
\max_{t'\in[0,t]}\int d^2\alpha\left(\prod_r\int d^2\beta_r\right)\left|c_0+c_a\alpha(t)+\sum_rc_r\beta_r(t)\right|^qQ_{N,t'}^{(0)},
\label{eq:A4_bound}
\eeq
in the time interval $0\leq t\leq t_0$.

\section{Upper bounds of $B_1'$ and $B_2'$}
\label{appendix:bound_B_prime}
\subsection*{Evaluation of $B_1'$}

First we consider the term $B_1'$.
Similarly to the case of $B_1$, we obtain the following expression:
\begin{align}
B_1'=&\sum_{l=0}^{L-1}\Bigg|\int_0^tdt_1\int_0^{t_1}dt_2\dots\int_0^{t_{l-1}}dt_l
\int d^2\alpha\left(\prod_r\int d^2\beta_r\right){\rm Tr}_{1,2,\dots,k+l}
\nonumber \\
&\quad\times\sum_{\{\sigma_j=\pm 1\}}^l X_{k+1}^{(\sigma_1)}X_{k+2}^{(\sigma_2)}\dots X_{k+l}^{(\sigma_l)}
\nonumber \\
&\quad\times
\frac{\d}{\d\alpha_{\sigma_l}(t_l)}\bigg\{\frac{\d}{\d\alpha_{\sigma_{l-1}}(t_{l-1})}\bigg\{
\dots \frac{\d}{\d\alpha_{\sigma_1}(t_1)}\bigg\{
\nonumber \\
&\quad\times{\cal O}^{(k)}(\alpha(t),\alpha^*(t))\tilde{U}_{t,t_1}^{(k)}\bigg\}
\tilde{U}_{t_1,t_2}^{(k+1)}\bigg\}\dots \tilde{U}_{t_{l-1},t_l}^{(k+l-1)}\bigg\}\tilde{U}_{t_l,0}^{(k+l)}
\left(Q_{N,0}^{(k+l)}-Q_{{\rm MF},0}^{(k+l)}\right)\Bigg|.
\end{align}
It is expressed as
\beq
B_1'=\sum_{l=0}^{L-1}\left|
\int d^2\alpha\left(\prod_r\int d^2\beta_r\right){\rm Tr}_{1,2,\dots,k+l}B_1^{(k+l)}
\left(Q_{N,0}^{(k+l)}-Q_{{\rm MF},0}^{(k+l)}\right)\right|,
\eeq
where $B_1^{(k+l)}$ was defined by Eq.~(\ref{eq:A1_k+l}).
From the assumption of the initial condition~(\ref{eq:initial}), we can conclude that $\lim_{N\rightarrow\infty}B_1'=0$
if we can show $B_1^{(k+l)}\in{\cal B}$.

In order to show $B_1^{(k+l)}\in{\cal B}$, we must consider the quantity 
\beq
\left\|\frac{\d^m B_1^{(k+l)}}{\d\alpha_{\tau_1}\d\alpha_{\tau_2}\dots\d\alpha_{\tau_m}}\right\|_{\infty},
\label{eq:A1_S}
\eeq
where $\tau_j=\pm 1$.

A calculation analogous to that for $B_1$ yields
\begin{align}
\left\|\frac{\d^m B_1^{(k+l)}}{\d\alpha_{\tau_1}\d\alpha_{\tau_2}\dots\d\alpha_{\tau_m}}\right\|_{\infty}
\leq\frac{(g\|X\|_{\infty}t)^l}{l!}\sum_{\{\sigma_j=\pm 1\}}\sum_{n=0}^l\sum_{N_{l,n}\in{\cal N}_{l,n}}
\sum_{p=0}^m\sum_{N_{m,p}\in{\cal N}_{m,p}}
\nonumber \\
\left\|\left(\prod_{j\in N_{l,n}}\frac{\d}{\d\alpha_{\sigma_j}(t)}\right)
\left(\prod_{j\in N_{m,p}^c}\frac{\d}{\d\alpha_{\tau_j}(t)}\right){\cal O}^{(k)}(\alpha(t),\alpha^*(t))\right\|_{\infty}
\nonumber \\
\times (k+l)^p(k+l-1)(k+l-2)\dots(k+n)\times(2g\|X\|_{\infty}t)^{l-n+p}.
\end{align}
Because ${\cal O}^{(k)}\in{\cal B}$, we have
\begin{align}
&\left\|\frac{\d^m B_1^{(k+l)}}{\d\alpha_{\tau_1}\d\alpha_{\tau_2}\dots\d\alpha_{\tau_m}}\right\|_{\infty}
\nonumber \\
&\leq\frac{(g\|X\|_{\infty}t)^l}{l!}\sum_{\{\sigma_j=\pm 1\}}\sum_{n=0}^l\sum_{N_{l,n}\in{\cal N}_{l,n}}
\sum_{p=0}^m\sum_{N_{m,p}\in{\cal N}_{m,p}}
\nonumber \\
&\times(k+l)^p(k+l-1)(k+l-2)\dots(k+n)(2g\|X\|_{\infty}t)^{l-n+p}
\nonumber \\
&\times\kappa^{n-m+p}\sum_{q=0}^sa_q\left|c_0+c_a\alpha(t)+\sum_rc_r\beta_r\right|^q.
\end{align}

Since $\sum_{\{\sigma_j=\pm 1\}}1=2^l$, $\sum_{N_{l,n}\in{\cal N}_{l,n}}1=l!/(n!(l-n)!)$, 
we obtain the upper bound
\begin{align}
&\left\|\frac{\d^m B_1^{(k+l)}}{\d\alpha_{\tau_1}\d\alpha_{\tau_2}\dots\d\alpha_{\tau_m}}\right\|_{\infty}
\nonumber \\
&\leq 2^{k+l-1}\sum_{n=0}^l\frac{1}{n!}\kappa^n(2g\|X\|_{\infty}t)^{2n-l}
\nonumber \\
&\times\left[\kappa+2(k+l)g\|X\|_{\infty}t\right]^m
\sum_{q=0}^sa_q\left|c_0+c_a\alpha(t)+\sum_rc_r\beta_r(t)\right|^q.
\label{eq:A1_diff_bound}
\end{align}
From this expression, we find that $B_1^{(k+l)}\in{\cal B}$ for any fixed $t$.
Thus we can conclude that
\beq
\lim_{N\rightarrow\infty}B_1'=0.
\eeq

\subsection*{Evaluation of $B_2'$}

$B_2'$ is given by
\begin{align}
B_2'=&\Bigg|\int d^2\alpha\left(\prod_r\int d^2\beta_r\right){\rm Tr}_{1,2,\dots,k}
{\cal O}^{(k)}\left(\alpha,\alpha^*,\{\beta_r,\beta_r^*\}\right)\times\int_0^tdt_1\int_0^{t_1}dt_2\dots\int_0^{t_{L-1}}dt_L
\nonumber \\
&\times U_{t,t_1}^{(k)}{\cal W}^{(k)}U_{t_1,t_2}^{(k+1)}{\cal W}^{(k+1)}
\dots U_{t_{L-1},t_L}^{(k+L-1)}{\cal W}^{(k+L-1)}\left(Q_{N,t_L}^{(k+L)}-Q_{{\rm MF},t_L}^{(k+L)}\right)\Bigg|
\end{align}
This is written in the form
\beq
B_2'\leq\int d^2\alpha\left(\prod_r\int d^2\beta_r\right){\rm Tr}_{1,2,\dots,k}\left\|B_2^{(k+L)}\right\|_{\infty}
\left(Q_{N,t_L}^{(k+L)}+Q_{{\rm MF},t_L}^{(k+L)}\right),
\label{eq:A2'}
\eeq
where $B_2^{(k+L)}$ is given by Eq.~(\ref{eq:A2_k+L}).
From the upper bound of $B_2^{(k+L)}$ (see Eq.~(\ref{eq:A2_k+L_bound})), it yields
\begin{align}
B_2'\leq 2^{-L+k-1}e^{2\kappa}
\int d^2\alpha\left(\prod_r\int d^2\beta_r\right){\rm Tr}_{1,2,\dots,k}\sum_{q=0}^sa_q\left|c_0+c_a\alpha(t)+\sum_rc_r\beta_r(t)\right|^q
\nonumber \\
\times\left(Q_{N,t_L}^{(k+L)}+Q_{{\rm MF},t_L}^{(k+L)}\right),
\label{eq:A2'_bound}
\end{align}
which is almost the same as Eq.~(\ref{eq:A2_bound}).
Because $$\lim_{N\rightarrow\infty}\int d^2\alpha\left(\prod_r\int d^2\beta_r\right)
{\rm Tr}_{1,2,\dots,k}\sum_{q=0}^s a_q\left|c_0+c_a\alpha(t)+\sum_rc_r\beta_r(t)\right|^q
\left(Q_{N,t_L}^{(k+L)}+Q_{{\rm MF},t_L}^{(k+L)}\right)$$
is finite and not diverging as $L\rightarrow\infty$, it is concluded that
\beq
\lim_{L\rightarrow\infty}\lim_{N\rightarrow\infty}B_2'=0,
\eeq
for $0\leq t\leq t_0$.

\end{document}